\documentclass[useAMS,usenatbib]{mn2e} 

\usepackage{amssymb,amsmath}
\usepackage{psfig,epsfig,amssymb,alltt}
\usepackage{abbrevjournaux}
\usepackage{latexsym}
\vspace{1cm}

\title[FASTLens (FAst STatistics for weak Lensing)]{FASTLens (FAst STatistics for weak Lensing) : Fast method for Weak Lensing Statistics and map making}
\author[S. Pires et al.]
  {S.~Pires,$^1$ J.-L.~Starck,$^1$ A.~Amara,$^1$$^,$$^2$
   R.~Teyssier,$^1$ A.~R\'efr\'egier, $^1$ J.~Fadili$^3$\\
  $^1$Laboratoire AIM, CEA/DSM-CNRS-Universite Paris Diderot, IRFU/SEDI-SAP, Service d'Astrophysique, \\CEA Saclay, Orme des Merisiers, 91191 Gif-sur-Yvette, France\\
  $^2$Department of Physics and Center for Theoretical and Computational Physics, The University of Hong Kong, \\Pok Fu Lam Road, Hong Kong\\
  $^3$GREYC CNRS UMR 6072, Image Processing Group, ENSICAEN 14050, Caen Cedex, France}
  \date{Released 2002 Xxxxx XX}

\pagerange{\pageref{firstpage}--\pageref{lastpage}} \pubyear{2002}

\def\LaTeX{L\kern-.36em\raise.3ex\hbox{a}\kern-.15em
    T\kern-.1667em\lower.7ex\hbox{E}\kern-.125emX}

\begin{document}

\label{firstpage}

\maketitle

\begin{abstract}
With increasingly large data sets, weak lensing measurements are able to measure cosmological parameters with ever greater precision. However this increased accuracy also places greater demands on the statistical tools used to extract the available information. To date, the majority of lensing analyses use the two point-statistics of the cosmic shear field. These can either be studied directly using the two-point correlation function, or in Fourier space, using the power spectrum.  But analyzing weak lensing data inevitably involves the masking out of regions for example to remove bright stars from the field. Masking out the stars is common practice but the gaps in the data need proper handling. 
In this paper, we show how an {\em inpainting} technique allows us to properly fill in these gaps with only $N \log N$ operations, leading to a new image from which we can compute straight forwardly and with a very good accuracy both the pow er spectrum and the bispectrum. We propose then a new method to compute the bispectrum with a polar FFT algorithm,
which has the main advantage of avoiding any interpolation in the Fourier domain.
Finally we propose a new method for dark matter mass map reconstruction  from shear observations which integrates this new inpainting concept.
A range of examples based on 3D N-body simulations illustrates the results.
\end{abstract}


\section{Introduction}
The distortion of the images of distant galaxies by gravitational lensing offers a direct way of probing the statistical properties of the dark matter distribution in the Universe; without making any assumption about the relation between dark and visible matter, see \cite*{wlens:bartelmann01,wlens:mellier99,wlens:Waerbeke01,wlens:mellier02,PSF:refregier03}. This weak lensing effect has been detected by several groups to derive constraints on cosmological parameters.  
Analyzing an image for weak lensing involves inevitably the masking out of regions to remove bright stars from the field. Masking out the stars is common practice but the gaps in the data need proper handling. 

At present, the majority of lensing analyses use the two point-statistics of the cosmic shear field. These can either be studied directly using the two-point correlation function
\citep{twopoint:maoli01,twopoint:refregier02,twopoint:bacon03,wlens:massey05}, or in Fourier space, using the power spectrum \citep{wlens:brown03}. 
Higher order statistical measures, such as three or four-point correlation functions have been studied \citep{threepoint:bernardeau03,threepoint:pen03,threepoint:jarvis03} and have shown to  provide additional constraints on cosmological parameters.

Direct measurement of the correlation function, through pair counting, is widely used since this method is not biased by missing data, for instance the ones arising from the masking of bright stars. However, this method is computationally intensive, requiring $O(N^2)$ operations. It is therefore not feasible to use it for future ultra-wide lensing surveys.  Measuring the power spectrum is significantly less demanding computationally, requiring $O(N \log N)$ operations, but is strongly affected by missing data. 
The estimation of power spectra from various types of data is becoming increasingly important also in other cosmological applications such as in the analysis of Cosmic Microwave Background (CMB) temperature maps or  galaxy clustering. In the literature, a large number of papers have appeared in the last few years that discuss various solutions 
to the problem of power spectrum estimation from large data sets with complete or 
missing data  \citep{ml:bond98,ml:ruhl03,ml:tegmark97,master:hivon02,master:hansen02,twopoint:szapudi01,hankel:szapudi01,ml:efstathiou04,twopoint:szapudi05}.
As explained in details in section~\ref{sect_ps_art}, they present however some  limitations  such as numerical instabilities which require to 
to regularize the solution. In this paper, we investigate an alternative approach based recent work in harmonic analysis

\subsection*{A new approach: inpainting}

Inpainting techniques are well known in the image processing litterature and are used to fill the gaps (i.e. to fill the missing data) by inferring a maximum information from the remaining data. In other words, it is an extrapolation of the missing information using some priors on the solution. We investigate in this paper how to fill-in judiciously  masked regions so as to reduce the impact of missing data on the estimation of the power spectrum and of higher order statistical measures. 
The inpainting approach we propose relies on a long-standing discipline in statistical estimation theory; estimation with missing data  \citep{inpainting:dempster77,inpainting:little87}. We propose to use an inpainting method that relies on the sparse representation of the data introduced by \cite{inpainting:elad05}. In this work, inpainting is stated as a linear inverse ill-posed problem, that is solved in a principled bayesian framework, and for which the popular Expectation-Maximization mechanism comes as a natural iterative algorithm because of physically missing data \citep{starck:jalal06}. Doing so, our algorithm exhibits the following advantages: it is fast, it allows to estimate any statistics of any order, the geometry of the mask does not imply any instability, the complexity of the algorithm does not depend on the mask nor on data weighting. We show that, for two different kinds of realistic mask (similar to that for CFHT and Subaru weak lensing analyses), we can reach an accuracy of about 1\% and  0.3\% on the power spectrum,  and an accuracy of about 3\% and 1\%  on the equilateral bispectrum. In addition, our method naturally handles more complicated inverse problems such as the estimation of the convergence map from masked shear maps.

This paper is organized as follows. Section~\ref{sect_sim_wldata} describes the simulated data that will be  used to validate the proposed methods, especially how large statistical samples of 3D N-body simulations of density distribution have been produced using a grid architecture and how 2D weak lensing mass maps have been derived. It also shows typical kinds of masks that need to be considered when analysing real data. Section~\ref{sect_stat} is an introduction to different statistics which are of interest in weak lensing data analysis and a fast and accurate algorithm to compute the equilateral bispectrum using a polar Fast Fourier Transform (FFT) is introduced. The speed of the bispectrum algorithm arises from the quickness of the polar Fourier transform and the accuracy comes from the output polar grid of the polar Fourier transform, thus avoiding the interpolation of coefficients in Fourier space. In section~\ref{sect_inpainting}, we present our inpainting method for gap filling in weak lensing data, and we show that it is a fast and accurate solution to the  missing data problem for second and third order statistics calculation. In section~\ref{sect_inp_kappa}, we propose a new approach to derive dark matter mass maps from incomplete weak lensing shear maps which uses the inpainting concept. Our conclusions are summarized in section~\ref{sect_ccl}.

\section{Simulations of weak lensing mass maps}
\label{sect_sim_wldata}

\subsection{3D N-body cosmological simulations}

We have run realistic simulated convergence mass maps derived from N-body cosmological simulations using the RAMSES code \citep{code:teyssier02}.
The cosmological model is taken to be in concordance with the $\Lambda$CDM model. We have chosen a model with the following parameters close to WMAP1 : $\Omega_m = 0.3$, $\sigma_8 = 0.9$, $\Omega_L = 0.7$, $h = 0.7$ and we have run 33 realizations of the same model. Each simulation has $256^3$ particles with a box size of 162 Mpc/h. We refined the base grid of $256^3$ cells when the local particle number exceeds 10. We further refined similarly each additional levels up to a maximum level of refinement of 6, corresponding to a spatial resolution of 10 kpc/h, making certain the particle shot noise remains at an acceptable level (see Fig. \ref{power9} and Fig. \ref{power9b}).

\subsection{Grid computing}

The simulation suite was deployed on a grid architecture designed under the project name {\it Grid'5000} \citep{grid:bolze06}.  For that purpose, we use a newly developed 
middle-ware called DIET \citep{diet:caron06} in order to compile and execute the RAMSES code on a widely inhomogeneous grid system. For this experiment \citep{lego:caniou07}, 12 clusters have been used on 7 sites for a duration time of 48 hours. In total, 816 grid nodes have been used for the present experiment, leading to the execution of 33 complete simulations. Note that this overall computation was completed within 2 days. This would have taken more than one month on a regular 32 processor cluster.

\subsection{2D Weak lensing mass map}
\label{2dmap}
In N-body simulations, which are commonly used in cosmology, the dark matter distribution is represented using discrete massive particles.  The simplest way to deal with these particles is to map their positions onto a pixelised grid.  In the case of multiple sheet weak lensing, we do this by taking slices through the 3D simulations.  These slices are then projected into 2D mass sheets.  The effective convergence can subsequently be calculated by stacking a set of these 2D mass sheets along the line of sight, using the lensing efficiency function. This is a procedure that has been used before by \cite{wlens:vale03}, where the effective 2D mass distribution $\kappa_e$ is calculated by integrating the density fluctuation along the line of sight. 
Using the Born approximation which neglects the facts that the light rays do not follow straight lines, the convergence (see eq. \ref{eq:kappa_convergence}) can be numerically expressed by  :
\begin{eqnarray}
\kappa_e \approx \frac{3H_0^2\Omega_m L}{2c^2}\sum_i\frac{\chi_i(\chi_0-\chi_i)}{\chi_0a(\chi_i)}\left(\frac{n_pR^2}{N_ts^2}-\Delta r_{f_i} \right)
\end{eqnarray}
where $H_0$ is the Hubble constant, $\Omega_m$ is the density of matter, $c$ is the speed of light, L is the length of the box $\chi$ are co-moving distances, with $\chi_0$ being the co-moving distance to the source galaxies. The summation is performed over the $i^{th}$ box. The number of particles associated with a pixel of the simulation is $n_p$, the total number of particles within a simulation is $N_t$ and $s=L_p/L$ where $L_p$ is the length of the plane doing the lensing. $R$ is the size of the 2D maps and $\Delta r{f_i} = \frac{r_2-r_1}{L}$ where $r_1$ and $r_2$ are co-moving distances.

\begin{figure}
\centerline{
\includegraphics[width=5.5cm, height=6.2cm]{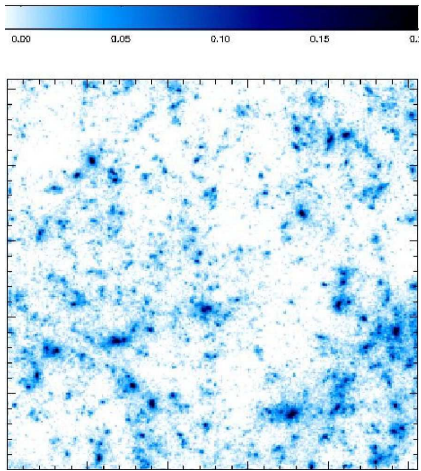}
}
\caption{Simulated weak lensing convergence map for the $\Lambda$CDM cosmological model with $\sigma_8 = 0.9$ and $\Omega_M = 0.3$. The region shown is $1^\circ$ x $1^\circ$.}
\label{model}
\end{figure}

We have derived $100$ simulated weak lensing mass maps from the previous 3D simulations.
Fig. \ref{model} shows a zoom of one of the 2D maps obtained by integration of the 3D density fluctuation on one of these realizations. The total field is $1.975 \times 1.975$ square degrees, with $512 \times 512$ pixels and we assume that the sources lie at exactly $z=1$. The overdensities (peaks) correspond to halos of groups and clusters of galaxies. The typical standard deviation values of $\kappa$ are thus of the order of a few percent.

\subsection{2D Weak lensing mass map with missing data}
\label{2dmaph}
Loss of data can be caused by many factors. Missing data can be due to camera CCD defect or from bright stars in the field of view that saturate the image around them as seen in weak lensing surveys with different telescopes \citep*{wlens:hoekstra06,wlens:hamana03,wlens:massey05,wlens:berge08}.

\begin{figure}
\centerline{
\hbox{
\includegraphics[width=3.cm, height=2.5cm]{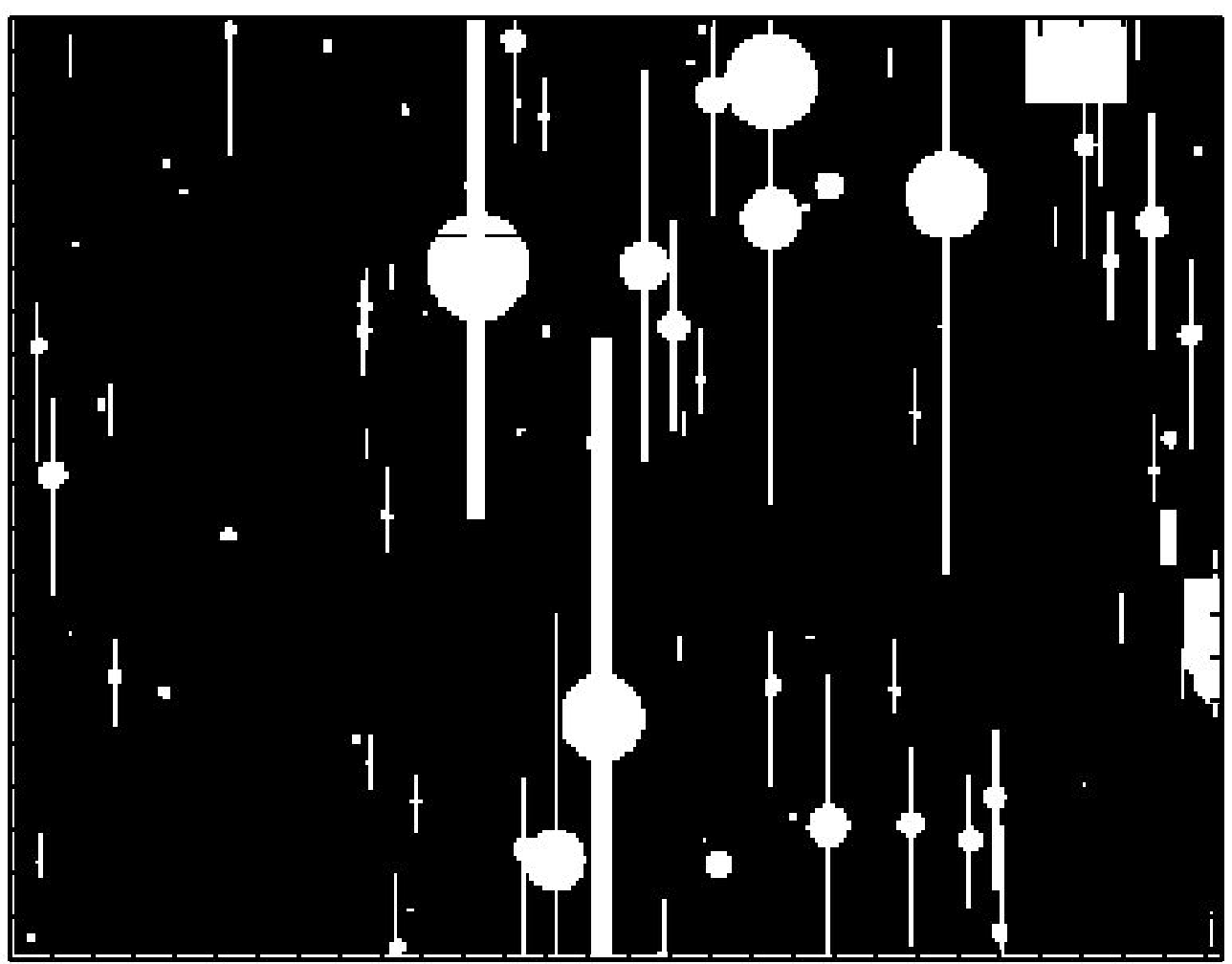}
\hspace{0.1cm}
\includegraphics[width=5.2cm, height=5.2cm]{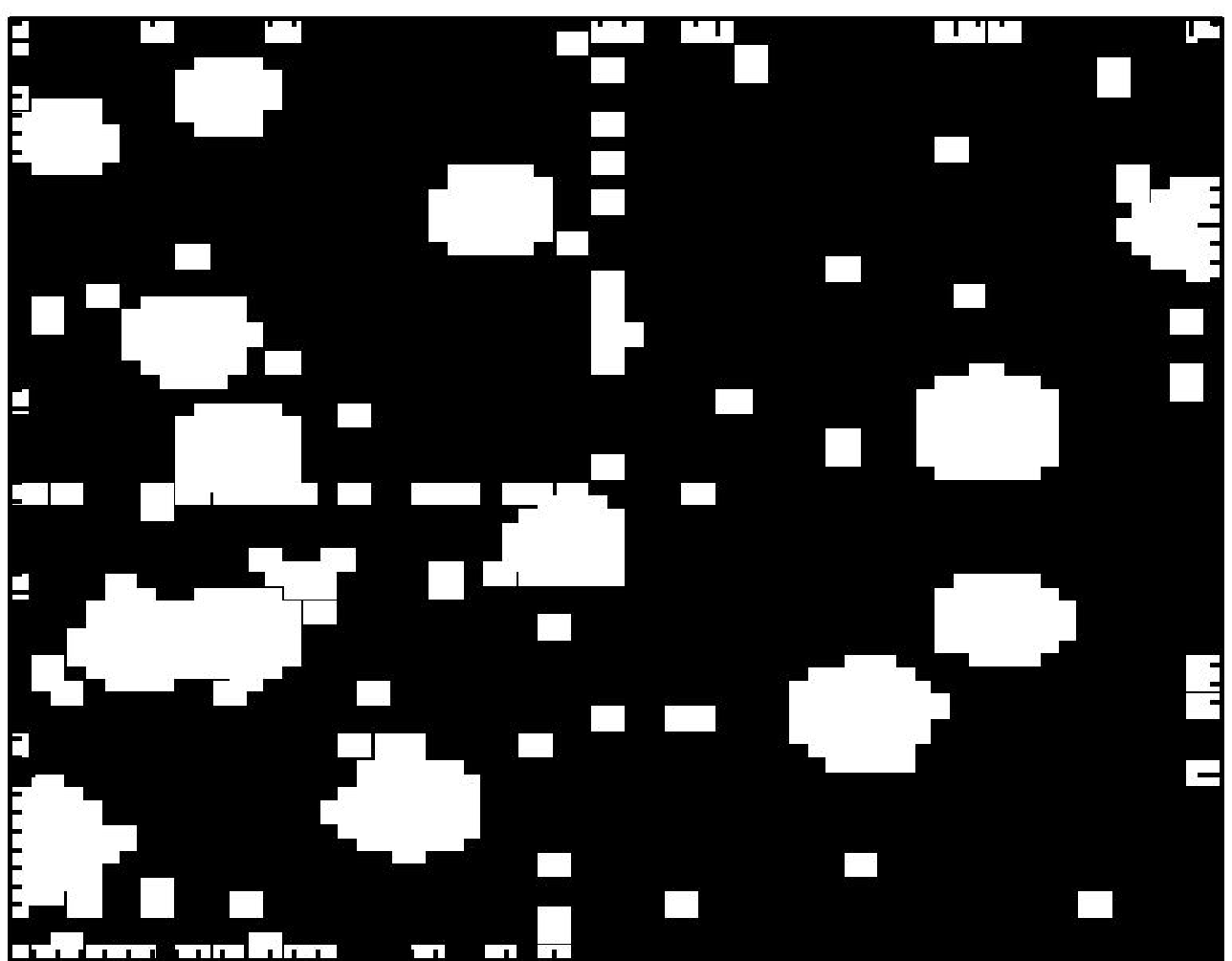}
}}
\caption{Left, mask pattern of Subaru Survey $0.575^\circ$ x $0.426^\circ$ (with SuprimeCam camera) right, mask pattern of CFHTLS data on D1 field $1^\circ$ x $1^\circ$ (with the MegaCam camera). Courtesy Joel Berge.}
\label{mask}
\end{figure}

Fig~\ref{mask} shows the mask pattern of CFHTLS image in the D1-field with about 20 $\%$ of missing data \citep{wlens:berge08} and that of SUBARU image covering a part of the same field with about 10 $\%$ of missing data. The mask pattern depends essentially on the field of view and on the quality of the optics.

In our simulations, we have chosen to consider these two typical 
mask patterns to study the impact of gaps in weak lensing analysis. The problem is now to extract statistical informations from weak lensing data with
such masks.

\section{Statistics}
\label{sect_stat}
\subsection{Two-point statistics}
\label{two-point}
Statistical weak gravitational lensing on large scales probes the projected density field of the matter in  the Universe : the convergence $\kappa$. Two-point statistics have become a standard way of quantifying the clustering of this weak lensing convergence field. Much of the interest in this type of analysis comes from its potential to constrain the spectrum of density fluctuations present in the late Universe. All second-order statistics of the convergence can be expressed as functions of the two-point correlation function of $\kappa$ or its Fourier transform, the Power Spectrum $P_{\kappa}$. 
\begin{itemize}

\item Two-point correlation function :\\
Direct two-point correlation function estimators $\tilde{C}_{\kappa \kappa}$ are based on the notion of pair counting. As a result of the Universe's statistical anisotropy, it only depends on $|\vec{\theta}|$ the distance between the position $\theta_i$ and the position $\theta_j$, and is given by :
\begin{eqnarray}
\tilde{C}_{\kappa \kappa}(|\theta_i - \theta_j|) = \frac{1}{N_{\theta}}  \sum^N_{i=1} \sum^N_{j=1}  \kappa(\theta_i) \kappa(\theta_j),
\label{eq:correl}
\end{eqnarray}
where $N_{\theta}$ is the number of pairs separated by a distance of $|\vec{\theta}|$.
Its na\"ive implementation requires $O(N^2)$ operations. Pair counting can be sped up, if we are interested in measuring the correlation function only on small scales. In that case the double-tree algorithm by \cite{fast:moore01} requires approximatively $O(N \log N)$ operations. However if all scales are considered, the tree-based algorithm slows down to $O(N^2)$ operations like na\"ive counting. 

\item Power spectrum :\\
The power spectrum $P_{\kappa}$ is the Fourier transform of the two-point correlation function (by the Wiener-Khinchine theorem). Because of the rotational invariance derived from the Universe isotropy, the Fourier transform becomes a Hankel transform :
\begin{eqnarray}
P_{\kappa}(q) &=& \frac{1}{2 \pi}\int_0^{+\infty}{C_{\kappa \kappa}(\theta) J_0(2 \pi q \theta) \theta d\theta},
\end{eqnarray}
where $J_0$ is the zero order Bessel function.
Computationally, we can estimate the power spectrum directly from the signal: 
\begin{eqnarray}
P_{\kappa}(q) &\propto& |\hat{\kappa}(q)|^2,
\end{eqnarray}
where $\hat{\kappa}$ denotes Fourier transform of the convergence.
Thus we can take advantage of the FFT algorithm to quickly estimate the power spectrum.

\item Sensitivity to missing data:\\
In weak lensing data analysis, it is common practice to mask out bright stars, which saturate the detector. This requires an appropriate post-treatment of the gaps. Contrarily to the two-point correlation function, the power spectrum estimation is strongly sensitive to missing data. Gaps generate a loss of power and gap edges produce distortions in the spectrum that depend on the size and the shape of the gaps.

\end{itemize}

\subsection{Three-point statistics}
\label{3pts}
Analogously with two-point statistics, third-order statistics are related to the three-point correlation function of $\kappa$ or its Fourier transform the Bispectrum $B_{\kappa}$.  
It is well established that the primordial density fluctuations are near Gaussian. Thus, the power spectrum alone contains all information about the large-scale structures in the linear regime. However, gravitational clustering is a non-linear process and in particular, on small scales, the mass distribution is highly non-gaussian. Three-point statistics are the lowest-order statistics to quantify non-gaussianity in the weak lensing field and thus provides additional information on structure formation models.

\begin{itemize}

\item Three-point correlation function :\\
Direct three-point correlation function estimators $C_{\kappa \kappa \kappa}$ are based on the notion of triangle counting. It  depends 
on  distances $d_1$, $d_2$ and $d_3$ between the three spatial positions $\theta_i$, $\theta_j$ and $\theta_k$  of  the triangle vertices formed by three galaxies, and is given by :
\begin{eqnarray}
C_{\kappa \kappa \kappa}(d_1, d_2, d_3) = <\kappa(\theta_i)\kappa(\theta_j) \kappa(\theta_k)>,
\label{eq:correl3_1}
\end{eqnarray}
where $< . >$  stands for the expected value.
The na\"ive implementation requires $O(N^3)$ operations and can consequently not be considered on future large data sets. 
One configuration, that is often used, is the equilateral configuration, wherein $d_1=d_2=d_3=d$. The three-point correlation function can then be plotted as a function of $d$. The configuration dependence being weak \citep{three-point:cooray01}, the equilateral configuration has become standard in weak lensing : first, because of its direct interpretation and second because its implementation is faster. 
The equilateral three-point correlation function estimation can be written as follows :
\begin{eqnarray}
\tilde{C}^{eq}_{\kappa \kappa \kappa}(d) = \frac{1}{N_d}  \sum^N_{i=1} \sum^N_{j=1} \sum^N_{k=1} \kappa(\theta_i) \kappa(\theta_j)  \kappa(\theta_k),
\label{eq:correl3_2}
\end{eqnarray}
where $N_{d}$ is the number of equilateral triangles whose side is $d$.
Whereas primary three-point correlation implementation requires $O(N^3)$ operations, equilateral triangle counting can be sped up to $O(N^2)$ operations. 
But this remains too slow to be used on future large data sets.

\item Bispectrum :\\
The complex Bispectrum is formally defined as the Fourier transform of the third-order correlation function. We assume the field $\kappa$ to be statistically isotropic, thus its bispectrum only depends on distances $|\vec{k_1}|$,  $|\vec{k_2}|$ and $|\vec{k_3}|$:
\begin{eqnarray}
B(|\vec{k_1}|, |\vec{k_2}|, |\vec{k_3}|) &\propto& <\hat{\kappa}(|\vec{k_1}|) \hat{\kappa}(|\vec{k_2}|) \hat{\kappa}^{*}(|\vec{k_3}|)>.
\label{eq:correl3_3}
\end{eqnarray}
If we consider the standard equilateral configuration, the triangles have to verify : $k_1=k_2=k_3=k$ and the bispectrum only depends on $k$ :
\begin{eqnarray}
B(k)^{eq} &\propto& <\hat{\kappa}(k) \hat{\kappa}(k) \hat{\kappa}^{*}(k)>.
\label{eq:correl3_4}
\end{eqnarray}

\item Sensitivity to missing data :\\
Like two-point statistics, the three-point correlation function is not biased by missing data. On  the contrary, the estimation of bispectrum is strongly sensitive to the missing data that produce important distortions in the bispectrum. For the time being, no correction has been proposed to deal with this missing data on bispectrum estimation and the three-point correlation function is computationally too slow to be used on future large data sets.
\end{itemize}

\subsection{The weak lensing statistics calculation from the polar FFT}
Assuming the gaps are correctly filled, the field becomes stationary and the Fourier modes uncorrelated. 
We present here a new method to calculate the power spectrum and the equilateral bispectrum accurately and efficiently. 

\begin{figure}
\centerline{
\includegraphics[width=6.5cm, height=6.cm]{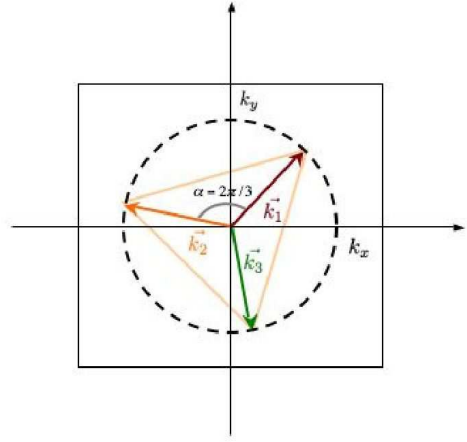}}
\caption{Equilateral bispectrum configuration in Fourier Space. Equilateral triangles must be inscribed in a circle of origin $(0,0)$ and of radius $k$. }
\label{fourier_config}
\end{figure}

To compute the bispectrum, we have to average over the equilateral triangles of length $k$ . 
Fig. \ref{fourier_config}  shows the form that such triangles in Fourier space. For each $k$, we integrate over all the equilateral triangles inscribed in the circle of origin $(0,0)$ and of radius $|\vec{k}|$. Because of the rotational symmetry we have only to scan orientation angles from $0$ to $2\pi/3$. The bispectrum is obtained by multiplying the Fourier coefficients located at the three vertices. 

Similarly, to compute the power spectrum, we have to average the modulus squared of the Fourier coefficients located in a circle of origin $(0,0)$ and of radius $k$.

\subsubsection*{The polar Fast Fourier transform (polar FFT)}
\begin{figure}
\centerline{
\includegraphics[width=3.7cm, height=3.7cm]{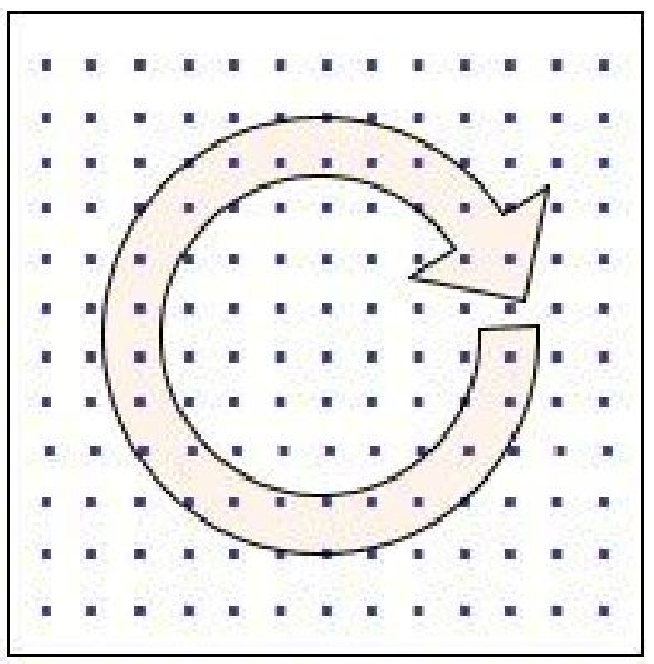}
\hspace{0.2cm}
\includegraphics[width=3.7cm, height=3.7cm]{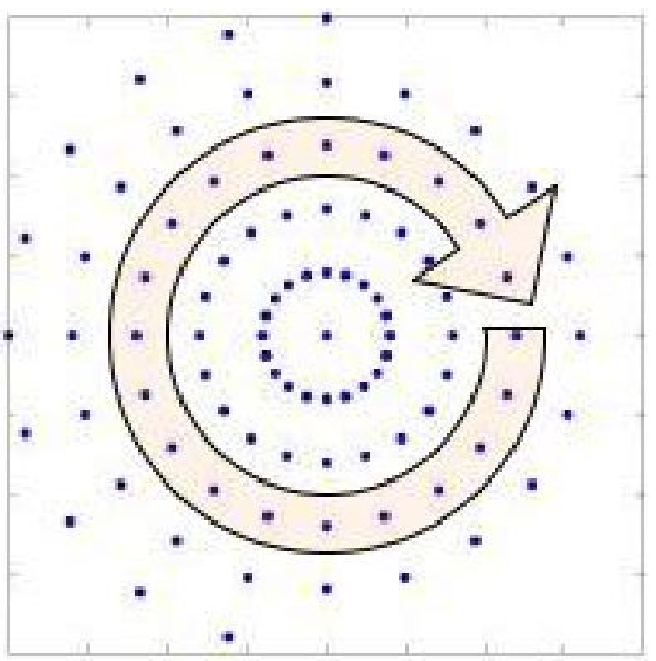}
}
\caption{Calculation of the mean power per frequency using a regular grid (left) and using a polar grid (right)}
\label{polar2}
\end{figure}

It requires some approximations to interpolate the Fourier coefficients in an equi-spaced Cartesian grid, as shown in Fig. \ref{polar2} on the left. In order to avoid these approximations, a solution consists in using a recent method, called polar Fast Fourier Transform that is a powerful tool to manipulate the Fourier transform in polar coordinates. 

A fast and accurate Polar FFT has been proposed by \cite{polar:averbuch05}.  For a given two-dimensional signal of size N, the proposed algorithm's complexity is $O(N \log N)$, just like in a Cartesian 2D-FFT. The polar FFT is just a particular case of the more general problem of finding the Fourier transform in a non-equispaced grid \citep{polar:keiner06}.  
We have used the NFFT (Non equi-spaced Fast Fourier Transform, software available at  {\em http://www-user.tu-chemnitz.de/$\sim$potts/nfft})  to compute a very accurate power spectrum and equilateral bispectrum.  Fig. \ref{polar2} (right) shows the grid that we have chosen. For each radius, we have the same number of points. The calculation of the average power associated to each equilateral triangle along the radius becomes easy and approximations are no longer needed.

\subsubsection*{Algorithms}
The Polar-FFT bispectrum algorithm is: 
\begin{center}
\begin{tabular}{|c|} \hline
\begin{minipage}[b]{8.cm}
\vspace{0.1in}
\small{
\textsf{1. Forward polar Fourier transform of the convergence $\kappa$.}

\textsf{2. Set the radius (in polar coordinates) $r$ to $0$. Iterate:}

\hspace{0.2cm} \textsf{3. Set the angle (in polar coordinates) $\theta$ to $0$. Iterate:}

\hspace{0.2cm} \hspace{0.2cm} \textsf{4. Locate the cyclic equilateral triangle whose one vertex}

\hspace{0.2cm} \hspace{0.2cm} \textsf{(or corner) have ($r$, $\theta$) as coordinate. A cyclic triangle is}

\hspace{0.2cm} \hspace{0.2cm} \textsf{a triangle inscribed in a circle it means the sides are chords}

\hspace{0.2cm} \hspace{0.2cm} \textsf{of the circle.}

\hspace{0.2cm} \hspace{0.2cm} \textsf{5. Perform the product of the Fourier coefficients located}

\hspace{0.2cm} \hspace{0.2cm} \textsf{at  the three corners of the cyclic equilateral triangle.}

\hspace{0.2cm} \hspace{0.2cm} \textsf{6. $\theta = \theta + \delta \theta$ and if $\theta < 2\pi/3$ return to step 4.}

\hspace{0.2cm} \textsf{7. Average the product over all the cyclic equilateral triangle}

\hspace{0.2cm} \textsf{inscribed in the circle of radius $r$.}

\hspace{0.2cm} \textsf{8. $r=r+1$ and if $r <  r_{max}$, return to Step 3.}
}
\vspace{0.05in}
\end{minipage}
\\\hline
\end{tabular}
\\ \vspace{0.1in}
\end{center}

The Fourier coefficients at the three corners of the cyclic equilateral triangle are easy to locate using a polar grid.
We don't need to interpolate to obtain the Fourier coefficient values as we have to do with a Cartesian grid. 
In addition to be accurate, this computation is also very fast. Indeed, in the simulated field that covers a region of $1.975^\circ \times 1.975^\circ$ with $512 \times 512$ pixels, using a 2.5 GHz processor PC-linux, about 60 seconds are needed to complete the calculation of the equilateral bispectrum and the process only requires $O(N \log N)$ operations. The bulk of computation is invested in the polar FFT calculation. This algorithm will be used in the experiments in section \S \ref{sect_inp_kappa}.\\

Similarly the Polar-FFT power spectrum algorithm is:
\begin{center}
\begin{tabular}{|c|} \hline
\begin{minipage}[b]{8cm}
\vspace{0.1in}
\small{
\textsf{1. Forward polar Fourier transform of the convergence $\kappa$.}

\textsf{2. Take the modulus squared of the polar Fourier transform of the convergence.}

\textsf{3. Set the radius (in polar coordinates) $r$ to $0$. Iterate:}

\textsf{4. Average the power over all the possible angles in the circle of radius $r$.}

 \textsf{5. $r=r+1$ and if $r <  r_{max}$, return to Step 4.}
}
\vspace{0.05in}
\end{minipage}
\\\hline
\end{tabular}
\\ \vspace{0.1in}
\end{center}

\subsection{The missing data problem: state of the art}
\label{sect_ps_art}
\subsubsection*{Second Order Statisitics}

 In the literature, a large number of studies have been presented that discuss various solutions to the problem of power spectrum estimation from large data sets with complete or missing data. These papers can be roughly grouped as follows: 
\begin{itemize}
\item Maximum likelihood (ML) estimator: two types of ML estimators have been discussed. The first uses an iterative Newton-type algorithm to maximize the likelihood score function without any assumed explicit form on the covariance matrix of the observed data \citep{ml:bond98,ml:ruhl03}. The second one is based on a model of the power-spectrum; see e.g. \cite{ml:tegmark97}. The ML framework allows to claim optimality in ML sense and to derive Cram\`er-Rao lower-bounds on the power spectrum estimate. However, ML estimators can become quickly computationally prohibitive for current large-scale datasets. Moreover, to correct for masked data, ML estimators involve a "deconvolution" (analogue to the PPS method below) that requires the inversion of an estimate of the Fisher information matrix. The latter depends on the mask and may be singular (semidefinite positive) as is the case for large galactic cuts, and a regularization may need to be applied.

\item Pseudo power-spectrum (PPS) estimators: \cite{master:hivon02} proposed the MASTER method for estimating power-spectra for both Cartesian and spherical grids. Their algorithm was designed to handle missing data such as the galactic cut in CMB data using apodization windows; 
see also \cite{master:hansen02}. These estimators can be evaluated efficiently using fast transforms such as the spherical harmonic transform for spherical data. Besides their fast implementation, PPS-based methods also allow us to derive an analytic covariance matrix of the power spectrum estimate under certain simplifying assumptions (e.g. diagonal noise matrix, symmetric beams,etc...). However, the deconvolution step in MASTER requires the invertion of a coupling matrix which depends on the power spectrum of the apodizing window. The singularity of this matrix strongly relies on the size and the shape of missing areas. Thus, for many mask geometries, the coupling matrix is likely to become singular, hence making the deconvolution step instable. To cope with this limitation, one may resort to regularized inverses. This is for instance the case in \cite{master:hivon02} where it is proposed to bin the CMB power spectrum. Doing so, the authors implicitly assume that the underlying power spectrum is piece-wise constant, which yields a loss of smoothness and resolution. 

A related class of sub-optimal estimators use fast evaluation of the two-point correlation function, which can then be transformed to give an estimate of the power spectrum. Methods of this type (used e.g. for the CMB) are described by \cite{twopoint:szapudi01,hankel:szapudi01,twopoint:szapudi05} (the SPICE method and its Euclidean version eSPICE). This class of estimators is closely related, though not exactly equivalent to the PPS estimator.
However, there are two issues with the estimation formulae given by \cite*{twopoint:szapudi01,twopoint:szapudi05}:

The first one concerns statistics. SPICE uses the Wiener-Khinchine theorem in order to compute the 2PCF in the direct space
by a simple division between the inverse Fourier transform of the (masked)  data power spectrum and the inverse Fourier transform of the mask power spectrum.
 But when data are masked or apodized, the resulting process is no longer wide-sense stationary and the Wiener-Khinchine  theorem is not strictly valid anymore.

The second one is methodological. Indeed, to correct for missing data, instead of inverting the coupling matrix in spherical harmonic or Fourier spaces as done in MASTER (the "deconvolution step"), \cite*{twopoint:szapudi05} suggest to invert the coupling matrix in pixel space. They accomplish this by dividing the estimated auto-correlation function of the raw data by that of the mask. But, this inversion (deconvolution) is a typical ill-posed inverse problem, and a direct division is unstable in general. This could be alleviated with a regularization scheme, which needs then to be specified for a given application.

MASTER has been designed for data on the sphere, and no code is available for Cartesian maps.
eSPICE has been proposed for computing the power spectrum of a set points (e.g. galaxies), and has not tested for maps where each pixel position has an associated value (i.e. weight). Therefore a public code for cartesian pixel maps remains to be developed.

\item Other estimators: some ML estimators that make use of the scanning geometry in a specific experiment were proposed in the literature. A hybrid estimator was also proposed that combines an ML estimator at low multipoles and PPS estimate at high multipoles \citep{ml:efstathiou04}.

\end{itemize}

The interested reader may refer to \cite{ml:efstathiou04} for an extended review, further details and a comprehensive comparative study of these estimators.

\subsubsection*{Third Order Statisitics}

The ML estimators discussed above heavily rely on the Gaussianity of the field, for which the second-order statistics are the natural and sufficient statistics. Therefore, to estimate higher order statistics (e.g. test whether the process contains a non-gaussian contribution or not), the strategy must be radically changed.  
Many authors have already addressed the problem of three-point statistics. In \cite{three-point:kilbinger05}, the authors calculate, from $\Lambda$CDM ray tracing simulations, third-order aperture mass statistics that contain information about the bispectrum. Many authors have already derived analytical predictions for the three-point correlation function and the bispectrum \citep[e.g.][]{three-point:ma00a,three-point:ma00b,three-point:scoccimarro01,three-point:cooray01}. Estimating three-point correlation function from data has already be done \citep{three-point:bernardeau02} but can not be considered in future large data sets because it is computationally too intensive. In the conclusion of \cite{twopoint:szapudi01}, the authors briefly suggested to use the $p$-point correlation functions with implementations that are at best $O(N(\log N)^{p-1})$. However, it was not clear if this suggestion is valid for the missing data case. \cite{three-point:scoccimarro98} proposed an algorithm to compute the bispectrum from numerical simulations using a Fast Fourier transform but without considering the case of incomplete data. This method is used by \cite{three-point:fosalba05} to estimate the bispectrum from numerical simulations in order to compare it with the analytic halo model predictions. Some recent studies on CMB real data \citep{three-point:komatsu05,three-point:yadav07} concentrate on the non-gaussian quadratic term of primordial fluctuations using a bispectrum analysis. 
But correcting for missing data the full higher-order Fourier statistics still remain an outstanding issue. In the next section, we propose an alternative approach, the inpainting technique, to derive 2nd order and 3rd order statistics and possibly higher-order statistics.

\section{Weak lensing mass map inpainting}
\label{sect_inpainting}
Here we describe a new approach, based on the inpainting concept. 

\subsection{Introduction}
In order to compute the different statistics described previously, we need to correctly take into account the missing data. 
We investigate here a new  approach to deal with the missing data problem in weak lensing data set, 
which is called {\em inpainting} in analogy with the recovery process museums experts use for old and deteriorated artwork. 
Inpainting techniques are well known in the image processing litterature and consist in filling the gaps (i.e. missing data).
In other words, it is an extrapolation of the missing information using some prior on the solution. For instance,
Guillermo Sapiro and his collaborators \citep{text:sapiro1,text:sapiro2,text:sapiro3}
use a prior relative to a smooth continuation of isophotes. This principle leads to nonlinear partial differential equation
(PDE) model, propagating information from the boundaries of the holes while guaranteeing smoothness in some way \citep{inpainting:chan01, Masnou02,Bornemann06,ChanNg06}. Recently, \cite{inpainting:elad05} introduced a novel inpainting algorithm that is capable of reconstructing both texture and smooth image contents. This algorithm is a direct extension of the MCA (Morphological Component Analysis), designed for the separation of an image into different semantic components \citep{Starck2004,inpainting:starck04}. The arguments supporting this method were borrowed from the theory of Compressed Sensing (CS) recently developed by \cite{Donoho04} and \cite{CandesRombergTao04,CandesTao05,CandesTao04}.
This new method uses a prior of sparsity in the solution. It assumes that there exists a dictionary (i.e. wavelet, Discrete Cosine Transform, etc) where the complete data is sparse and where the incomplete data is less sparse. For example, mask borders are not well represented
in the Fourier domain and create many spurious frequencies, thus minimizing the number of frequencies is a way to enforce the sparsity in the Fourier dictionary. 
The solution that is proposed in this section is to judiciously fill-in masked regions so as to reduce the impact of missing data on the estimation of the power spectrum and of higher order statistical measures. 

\subsection{Inpainting based on sparse decomposition}

The classical image inpainting problem can be defined as follows. 
Let $X$ be the ideal complete image, $Y$ the observed incomplete image and $M$ the binary mask (i.e. $M_i = 1$ if we 
have information at pixel $i$, $M_i = 0$ otherwise). In short, we have: $Y = M X$.  
Inpainting consists in recovering $X$ knowing $Y$ and $M$.
 
In many applications - such as compression, de-noising, source separation and, of course, inpainting - a good and efficient
 signal representation is necessary to improve the quality of the processing. All representations are not equally interesting and there is a strong a priori for sparse representation because it makes information more concise and possibly more interpretable. 
This means that we seek a representation $\alpha = \Phi^T X$ of the signal $X$ in the dictionary $\Phi$ 
where most coefficients $\alpha_i$ are close to zero, while only a few have a significant absolute value.

Over the past decade, traditional signal representations have been replaced by a large number of new multiresolution representations.
Instead of representing signals as a superposition of sinusoids using classical Fourier representation, we now have many available alternative dictionaries such as wavelets \citep{wave:mallat89}, ridgelets \citep{ridge:candes99} or curvelets 
\citep{starck:sta02_3,cur:candes06}, most of which are overcomplete. This means that some elements of the dictionary can be described in terms of other ones, therefore a signal decomposition in such a dictionary is not unique. Although this can increase the complexity of the signal analysis, it gives us the possibility to select among many possible representations the one which gives the sparsest representation of our data.
 
To find a sparse representation and noting $|| z ||_0$ the $l_0$ pseudo-norm, i.e. the number of non-zero entries in $z$ and $|| z ||$ the classical $l_2$ norm (i.e. $ || z ||^2 = \sum_k (z_k)^2 $), we want to minimize:
\begin{equation}
\min_X  \| \Phi^T X \|_0 \quad   \text{subject to}  \quad   \parallel Y - M X  \parallel^2 \le \sigma,
\label{minimisation}
\end{equation}
where $\sigma$ stands for the noise standard deviation in the noisy case. Here, we will assume that no noise perturbs the data $Y$, 
$\sigma=0$ (i.e. the constraint becomes an equality). As discussed later, extension of the method to deal with noise is straighforward.

It has also been shown that if $\Phi^T X$ is sparse enough, the $l_0$ pseudo-norm can also be replaced by the convex $l_1$ norm (i.e. $ || z ||_1 = \sum_k | z_k | $) \citep{cur:donoho_01b}. The solution of such an optimisation task can be obtained through an iterative thresholding algorithm called MCA \citep{inpainting:elad05} :
\begin{equation}
   X^{n+1} = \Delta_{\Phi,\lambda_n}(X^{n} + M(Y - X^n)),
\label{eqn_mca}
\end{equation}
where the nonlinear operator $\Delta_{\Phi,\lambda}(Z)$ consists in:
\begin{itemize}
\item decomposing the signal $Z$ on the dictionary $\Phi$ to derive the coefficients $\alpha = \Phi^T Z$.
\item threshold the coefficients: ${\tilde \alpha} = \rho(\alpha, \lambda)$, 
where the thresholding operator $\rho$   can either
be a hard thresholding (i.e. $\rho(\alpha_i, \lambda) = \alpha_i$ if $ | \alpha_i | > \lambda$ and $0$ otherwise)
 or a soft thresholding (i.e.
  $\rho(\alpha_i, \lambda) = \mathrm{sign}(\alpha_i) \mathrm{max}(0, | \alpha_i |  - \lambda)$). 
 The hard thresholding corresponds to the $l_0$ 
 optimization problem while the soft-threshold solves that for $l_1$.
\item reconstruct $\tilde Z$ from the thresholded coefficients ${\tilde \alpha}$.
\end{itemize}
The threshold parameter $\lambda_n$ decreases with the iteration number and it plays a part similar to the cooling parameter  
of the simulated annealing techniques, i.e. it allows the solution to escape from local minima. 
More details relative to this optimization problem can be found in \cite{CombettesWajs05,starck:jalal06}. For many dictionaries such as wavelets
or Fourier, fast operators exist to decompose the signal so that the iteration of eq.~\ref{eqn_mca} is fast. It requires us only to perform,
at each iteration, a forward transform, a thresholding of the coefficients and an inverse transform. The case where the 
dictionary is a union of subdictionaries $\Phi = \{\Phi_1, ..., \Phi_T\}$ where each $\Phi_i$ has a fast operator has also been
investigated in \cite{inpainting:starck04,inpainting:elad05,starck:jalal06}.
We will discuss in the following the choice of the dictionary for the weak lensing inpainting problem.

In general, there are some restrictions in the use of inpainting based on sparsity that arise from the link between the sparse representation dictionary and the masking operator M. The first restriction is that the proposed inpainting method based on sparse representation assumes that a good representation for the data is not a good representation of the gaps. This means that features of the data need few coefficients to be represented, but if a gap is inserted in the data a large number of coefficients will be necessary to account for this gap. Then by minimizing the 
number of coefficients among all the possible coefficients, the initial data can be approximated.
 Secondly, the inpainting is possible if the gaps are smaller than the largest dictionary elements. Indeed, if a gap removes a part of an object that is well represented by one element of the dictionary, this object can be recovered. Obviously, if the whole object is missing, it can not be recovered.

\subsection{Sparse representation of weak lensing mass maps}

Representing the image to be inpainted in an appropriate sparsifying dictionary is the main issue. The better the dictionary, the better the inpainting quality is. We are interested in a large and overcomplete dictionary that can be also built by the union of several sub-dictionaries, each of which must be particularly suitable for describing a certain feature of a structured signal. For computational cost considerations, we consider only (sub-) dictionaries associated to fast operators.
Finding a sparse representation for weak lensing analysis is challenging. 
We want to describe well all the features contained in the data. 
The weak lensing signal is composed of clumpy structures such as clusters and filamentary structures (see Fig. \ref{model}). The weak lensing mass maps thus exhibit both isotropic and anisotropic features. The basis that best represent isotropic objects are not the same as those that best represent anisotropic ones. We have consequently investigated a number of sub-dictionaries and various combinations of sub-dictionaries :\\
- isotropic sub-dictionaries such as the ``\`a trous" wavelet representation \\
- slightly anisotropic sub-dictionaries such as the bi-orthognal wavelet representation \\
- highly anisotropic sub-dictionaries such as the curvelet representation \\
- texture sub-dictionaries such as the Discrete Cosine Transform \\

\begin{figure}
\centerline{
\includegraphics[width=7.5cm, height=6.5cm]{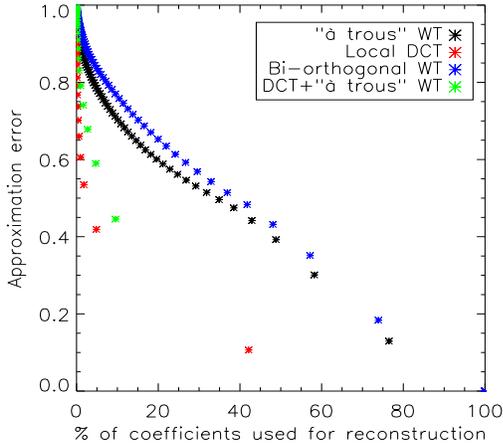}
}
\caption{Non-linear approximation error $l_2$ as a function of the percentage of coefficients used for the reconstruction, obtained with (i) the ``\`a trous" Wavelet Transform (black), (ii) the local Discrete Cosine Transform (with a blocksize of 256 pixels) (red), (iii) the bi-orthognal Wavelet Transform (blue) and (iv) the ``\`a trous" Wavelet Transform + the local DCT (blocksize = 256 pixels) (green). The better representation for weak lensing data is obtained with a local DCT.}
\label{l2error}
\end{figure}

A simple way to test the sparsity of a representation consist of estimating the non-linear approximation error $l_2$ from complete data. It means, the error obtained by keeping only the N largest coefficients in the inverse reconstruction. Fig. \ref{l2error} shows the reconstruction error $l_2$ as a function of N.\\

We expected wavelets to be the best dictionary because they are good to represent isotropic structures like clusters but surprisingly the better representation for weak lensing data is obtained with the DCT.
 More sophisticated representations recover clusters well, but neglect the weak lensing texture. Even combinations of DCT with others dictionaries (isotropic or not) is less competitive. We therefore chose DCT in the rest of our analysis.

\subsection{Algorithm}
\label{algorithm1}
Our final dictionary $\Phi$ being chosen, we want to minimize the number of non-zero coefficients (see eq. \ref{minimisation}).
We use an iterative algorithm for image inpainting as in \cite{inpainting:elad05} that we describe below.
This algorithm needs as inputs the incomplete image $Y$ and the binary mask $M$. The algorithm that we implement is:
\begin{center}
\begin{tabular}{|c|} \hline
\begin{minipage}[b]{8.cm}
\vspace{0.1in}
\small{
\textsf{1. Set the maximum number of iterations $I_{max}$, the solution $X^{0}=0$, the residual
$R^{0}$ to $Y$, the maximum threshold $\lambda_{max} = \mathrm{\max}(|\alpha = \Phi^T Y|$), the minimum threshold
$\lambda_{min}=0$.}

\textsf{2. Set $n$ to $0$, $\lambda_n = \lambda_{max}$. Iterate:}

\hspace{0.2cm} \textsf{3. $U = X^n + M R^n$ and $R^n = (Y - X^n)$.}

\hspace{0.2cm} \textsf{4. Forward transform of $U$: $\alpha = \Phi^T U$.}

\hspace{0.2cm} \textsf{5. Determination of the threshold level}

\hspace{0.2cm} \textsf{$\lambda_n = F(n, \lambda_{max}, \lambda_{min})$.}

\hspace{0.2cm} \textsf{6. Hard-threshold the coefficient $\alpha$ using $\lambda_n$ : $\tilde{\alpha}=S_{\lambda_n}\{\alpha\}$.}

\hspace{0.2cm} \textsf{7. Reconstruct $\tilde U$ from thresholded $\alpha$ and $X^{n+1}=\Phi \tilde{\alpha}$.}

\hspace{0.2cm} \textsf{8. $n=n+1$ and if $n <  I_{max}$, return to Step 3.}
}
\vspace{0.05in}
\end{minipage}
\\\hline
\end{tabular}
\\ \vspace{0.1in}
\end{center}

$\Phi^T $ is the DCT operator. The way the threshold is decreased at each step is important. It is a trade-off between the speed of the algorithm and its quality. The function $F$ fixes the decreasing of the threshold. A linear decrease corresponds to 
$F(n, \lambda_{max}, \lambda_{min}) = \lambda_{max} - \frac{n(\lambda_{max} - \lambda_{min})}{I_{max}-1}$. 

In practice, we use a faster decreasing law defined by: 
$F(n, \lambda_{max}, \lambda_{min}) =  \lambda_{min}  + (\lambda_{max} - \lambda_{min})  (1.-\mathrm{erf}(2.8 n/ I_{max}))$.

Constraints can also be added to the solution. For instance, we can, at each iteration, enforce the variance of the solution $X^{n+1}$ to be equal inside and outside the masked region. We found that it improves the solution.

\begin{figure}
\centerline{
\includegraphics[width=7.5cm, height=5.cm]{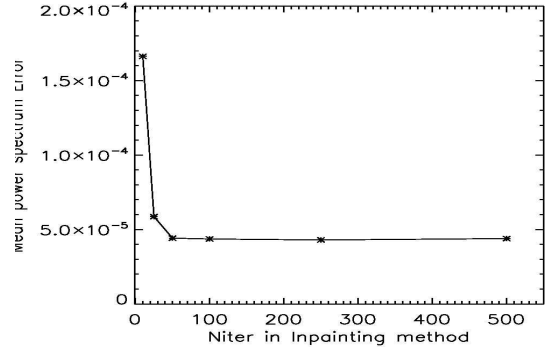}
}
\caption{Mean power spectrum error as a function of the maximum number of iteration used by our inpainting method with the CFHTLS mask.}
\label{iter}
\end{figure}

The only parameter is the number of iterations $I_{max}$. In order to see the impact of this parameter, we made the following experiment : we estimated the mean power spectrum error $<E_{P_{\kappa}}(I_{max})>$ for different values of $I_{max}$, see Fig. \ref{iter}. 
The mean power spectrum error is defined as follows:
\begin{eqnarray}
<E_{P_{\kappa}}(I_{max})>=\frac{1}{N_m} \sum_m \left[ \frac{1}{N_q} \sum_q (P_{\tilde{\kappa}_{I_{max}}^m}(q) - P_{\kappa}(q))^2\right],
\label{MPSE}
\end{eqnarray}

where $\tilde{\kappa}_{I_{max}}^m$ stands for the $m^{th}$ inpainted map with $I_{max}$ iterations, $N_q$ is the number of bins in the power spectrum and $N_m$ is the number of maps over which we estimate the mean power spectrum.

 It is clear that the error on the power spectrum decreases and reaches a plateau for $I_{max} > 100$. We thus set the number of iterations to 100.

\subsection{Handling noise}
\label{noise}
If the data contains noise, it is straightforward to take it into account. Indeed, thresholding techniques are
very robust to the noise since they are even used to remove it. In the preceding algorithm, a filtered inpainted image
can directly be obtained by setting the final threshold $\lambda_{min}$ to $\tau \sigma$ instead of $0$, where $\sigma$ is the 
noise standard deviation and $\tau$ is a constant generally choosen between 3 and 5. 
However, even if the data is noisy, we may not want to perform denoising because it could introduce a bias in a further
analysis such as power spectrum analysis. In this case, the final threshold $\lambda_{min}$ should be kept at $0$ and 
the inpainting will try to reproduce also the noise texture (as if it were a real signal). Obviously, it won't recover the 
"true" noise that will be observed if there was no missing data, but the statistical properties should be similar to that in the
rest of the image. As we will see in the following, our experiments confirm this assertion.

\subsection{Experiments}

We present here several experiments to show how we have lowered the impact of the mask by applying the MCA-inpainting algorithm described above. The MCA-inpainting algorithm was applied with 100 iterations and using the DCT representation over the set of 100 incomplete mass maps, as described previously. The case of noisy mass maps has not been considered in the following experiments, it will be studied in \S \ref{inp_shear}.

\subsubsection*{Incomplete mass map interpolation by MCA-inpainting algorithm}
\begin{figure*}
\vbox{
\centerline{
\hbox{
\includegraphics[width=6.5cm, height=6.5cm]{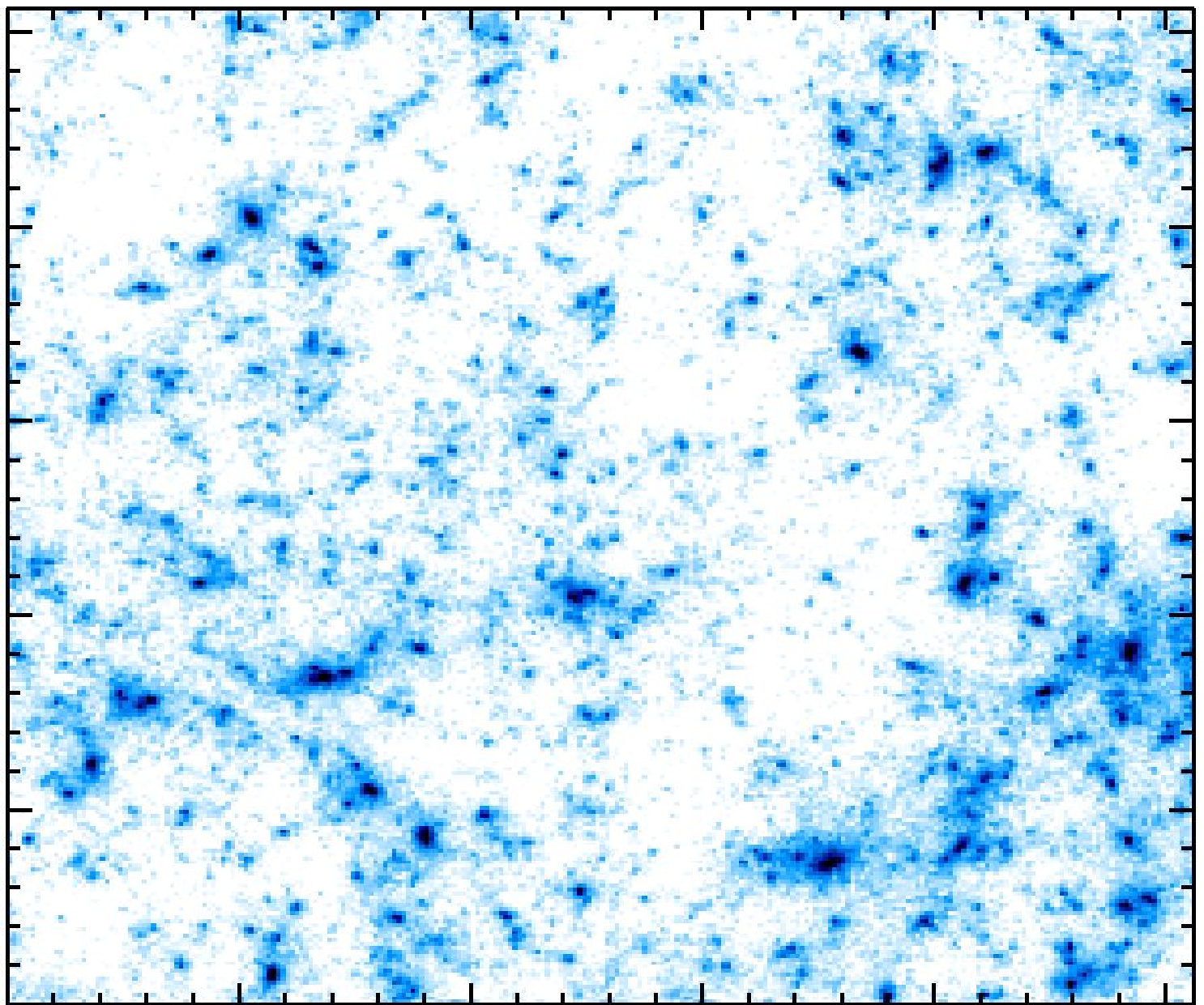}
\hspace{0.2cm}
\includegraphics[width=6.5cm, height=6.5cm]{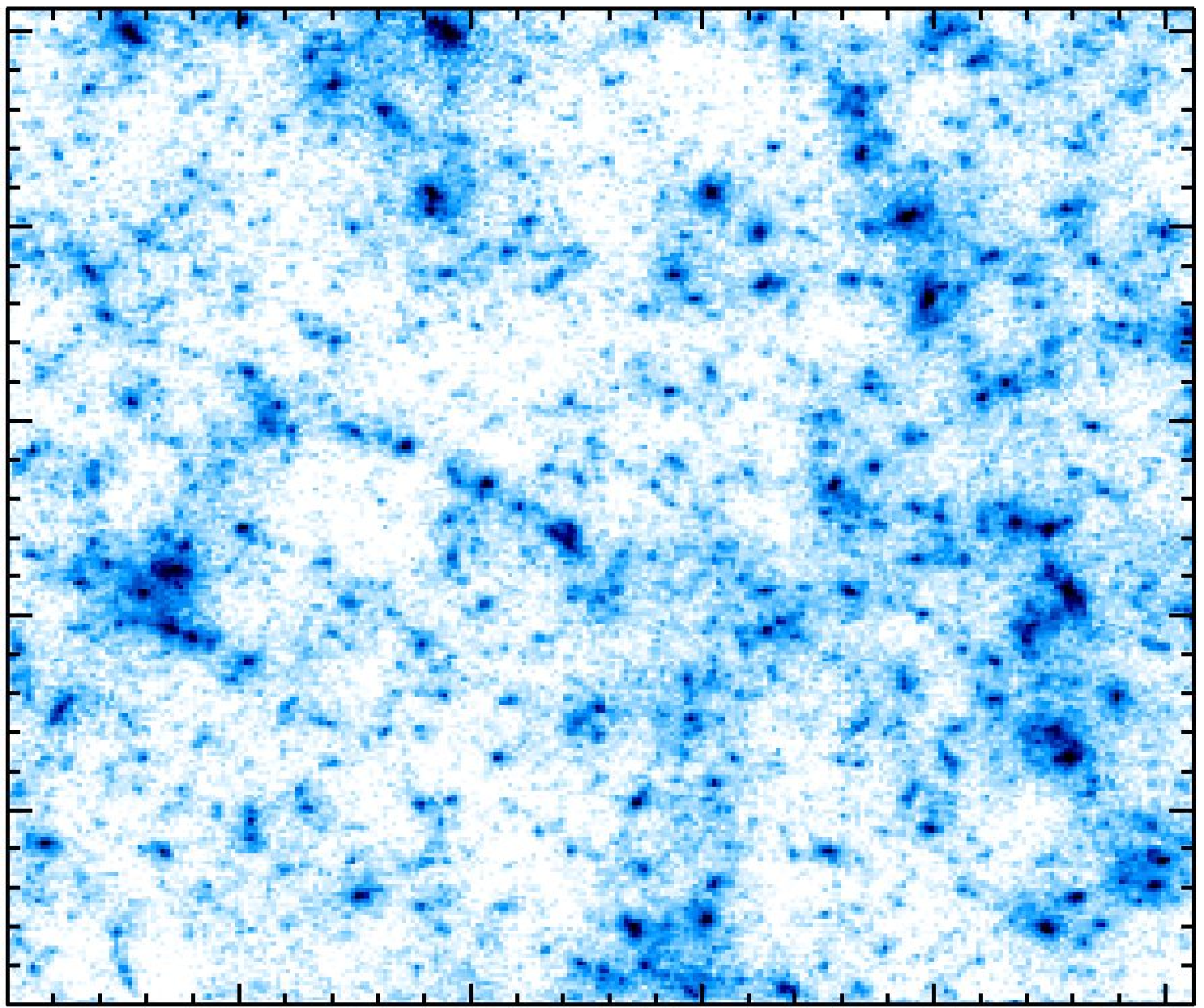}
}}
\vspace{0.2cm}
\centerline{
\hbox{
\includegraphics[width=6.5cm, height=6.5cm]{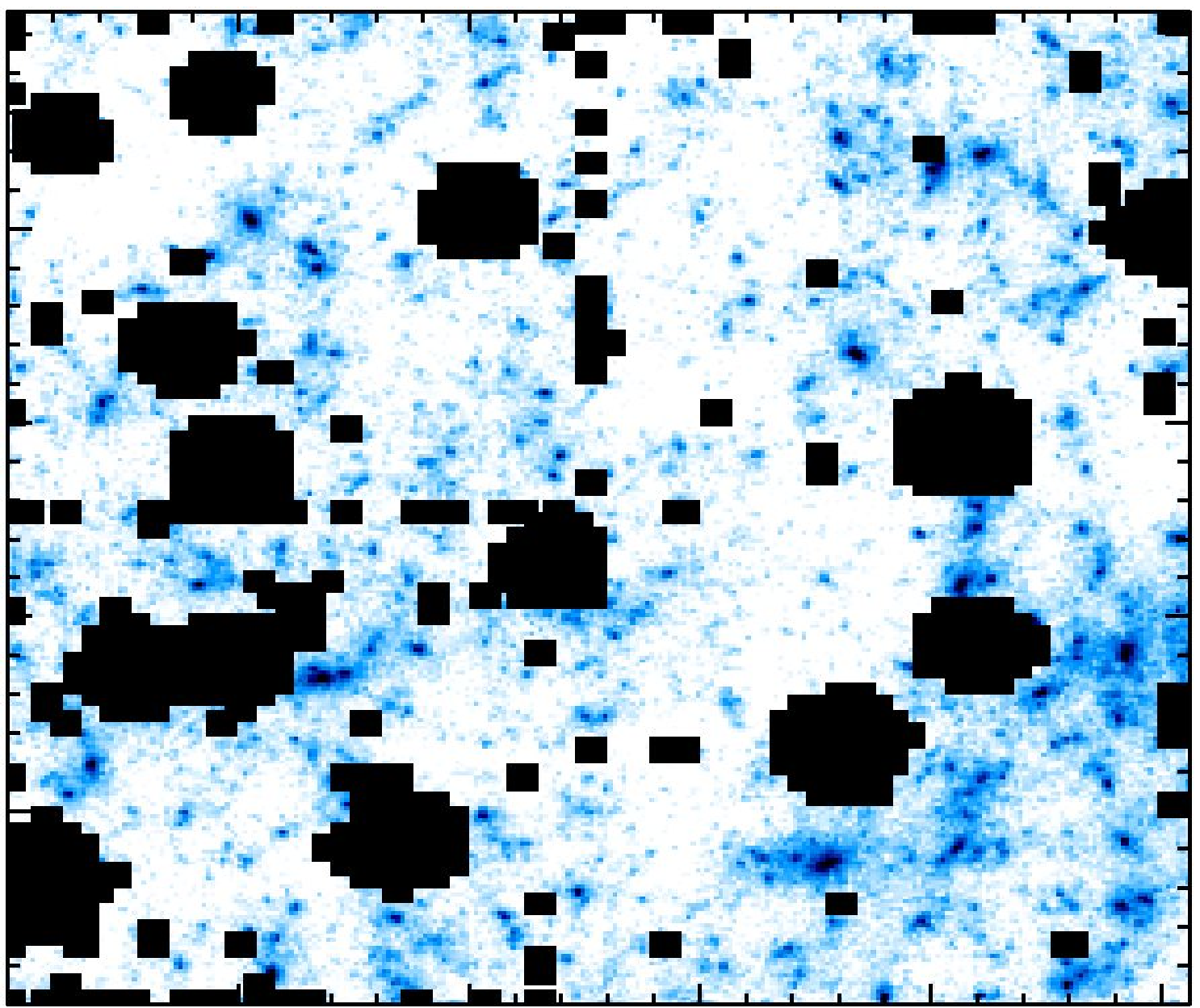}
\hspace{0.2cm}
\includegraphics[width=6.5cm, height=6.5cm]{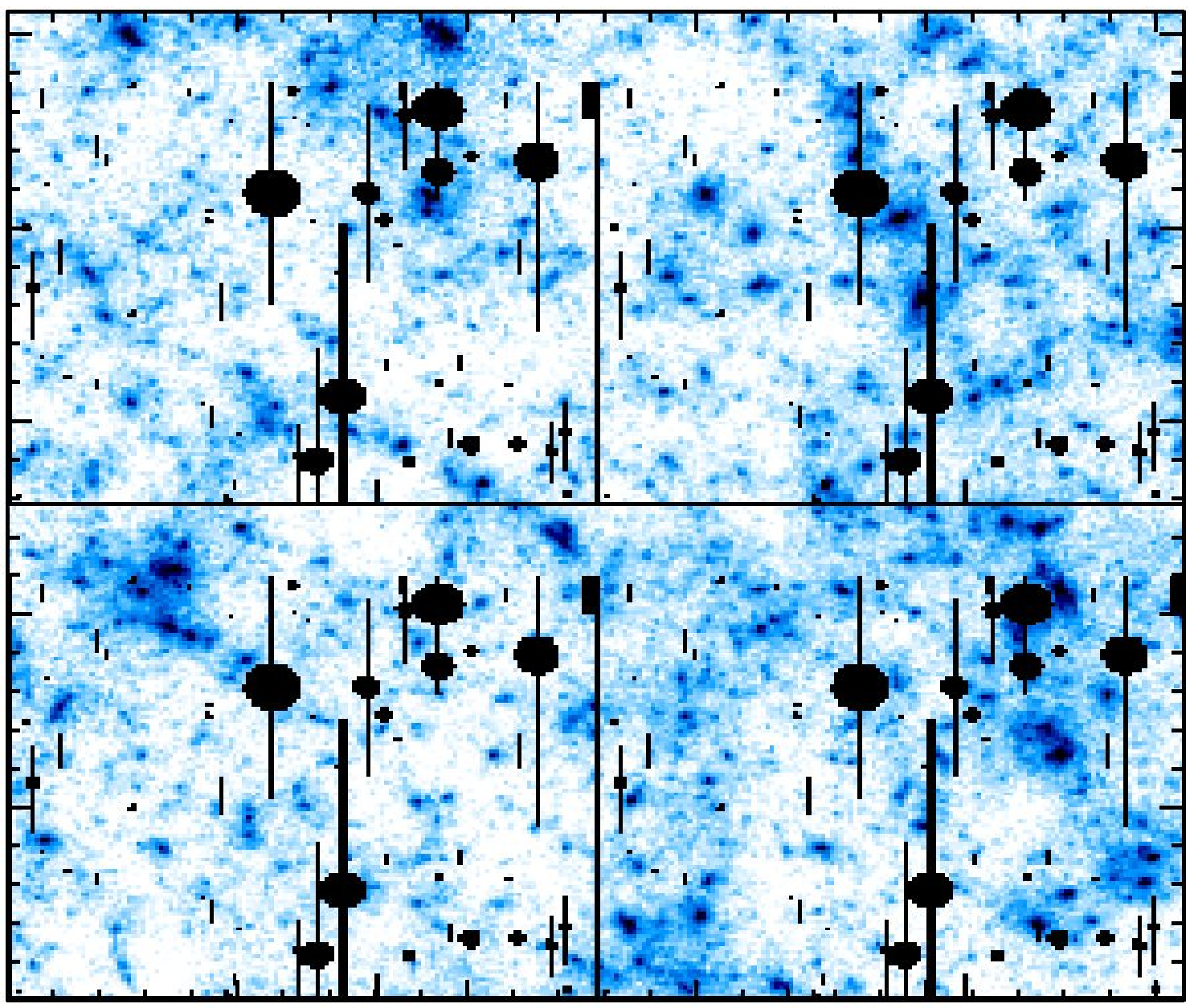}
}}
\vspace{0.2cm}
\centerline{
\hbox{
\includegraphics[width=6.5cm, height=6.5cm]{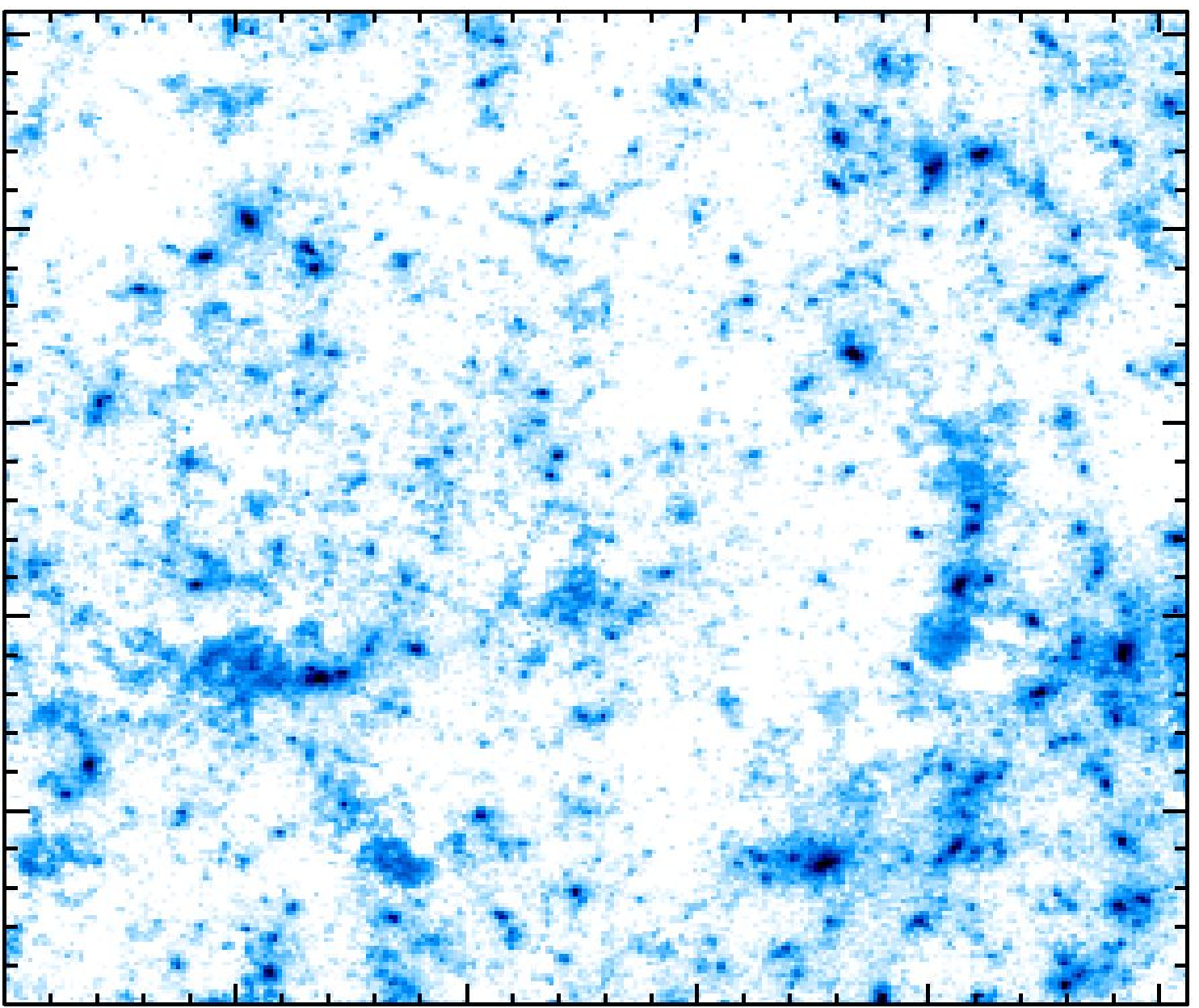}
\hspace{0.2cm}
\includegraphics[width=6.5cm, height=6.5cm]{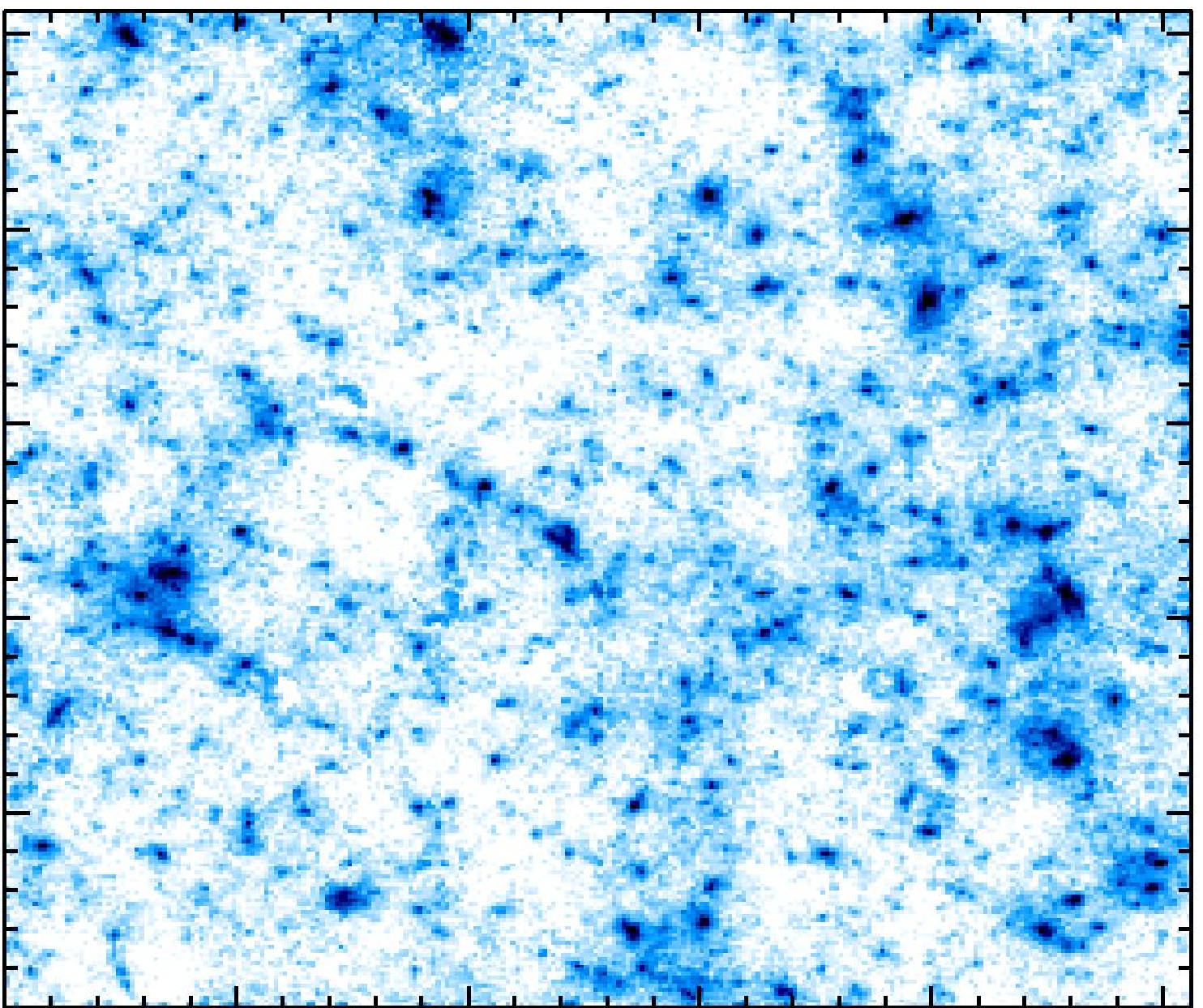}
}}
}
\caption{Upper panels, simulated weak lensing mass map, middle panels, simulated mass map with the mask pattern of CFHTLS data on D1 field (left) and with the mask pattern of Subaru data in the same field (right), lower panels, inpainted mass map. The region shown is $1^\circ$ x $1^\circ$.}
\label{inpainting}
\end{figure*}

The first experiment was conducted on two different simulated weak lensing mass maps masked by two typical mask patterns (see Fig. \ref{inpainting}). 
The upper panels show the simulated mass maps (see \S \ref{2dmap}), the middle panels show the simulated mass map masked by the CFHTLS mask (on the left) and  the Subaru mask (on the right) (see \S \ref{2dmaph}). The results of the MCA-inpainting method using the DCT decomposition is shown in the lower panels allowing a first visual estimation of the quality of the proposed algorithm. We note that the gaps are no longer distinguishable by eye in the inpainted map.

 \subsubsection*{Probability Density Function comparison}

\begin{figure}
\centerline{
\includegraphics[width=8.cm, height=6.cm]{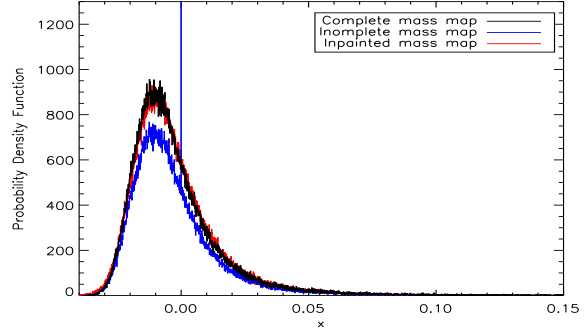} 
}
\caption{Probability Density Function estimated from the 3 maps on the left of the Fig.\ref{inpainting} : PDF of the complete simulated mass map (in black), PDF of the incomplete mass map (in blue) and the PDF of the inpainted mass map (in red).}
\label{pdf}
\end{figure}

This second experiment was conducted on the weak lensing mass maps represented Fig.\ref{inpainting} on the left. 
For these maps, we have computed the Probability Density Function (PDF) in order to compare their statistical distributions (see Fig. \ref{pdf}).
The PDF of the complete mass map is plotted as a solid black line, the PDF of the incomplete mass map as a solid blue line and the PDF of the inpainted mass map as a solid red line. The blue vertical line corresponds to the pixels masked out in incomplete mass maps. 
The strength of the PDF is that it provides a visual estimation of some statistics like the mean, the standard deviation, the skewness and the kurtosis. 
Thus, it provides a way to directly quantify the quality of the inpainting method. 
A visual comparison shows a striking similarity between the inpainted distribution and the original one.

\subsubsection*{Power spectrum estimation}

\begin{figure*}
\vbox{
\centerline{
\hbox{
\includegraphics[width=8.cm, height=5.cm]{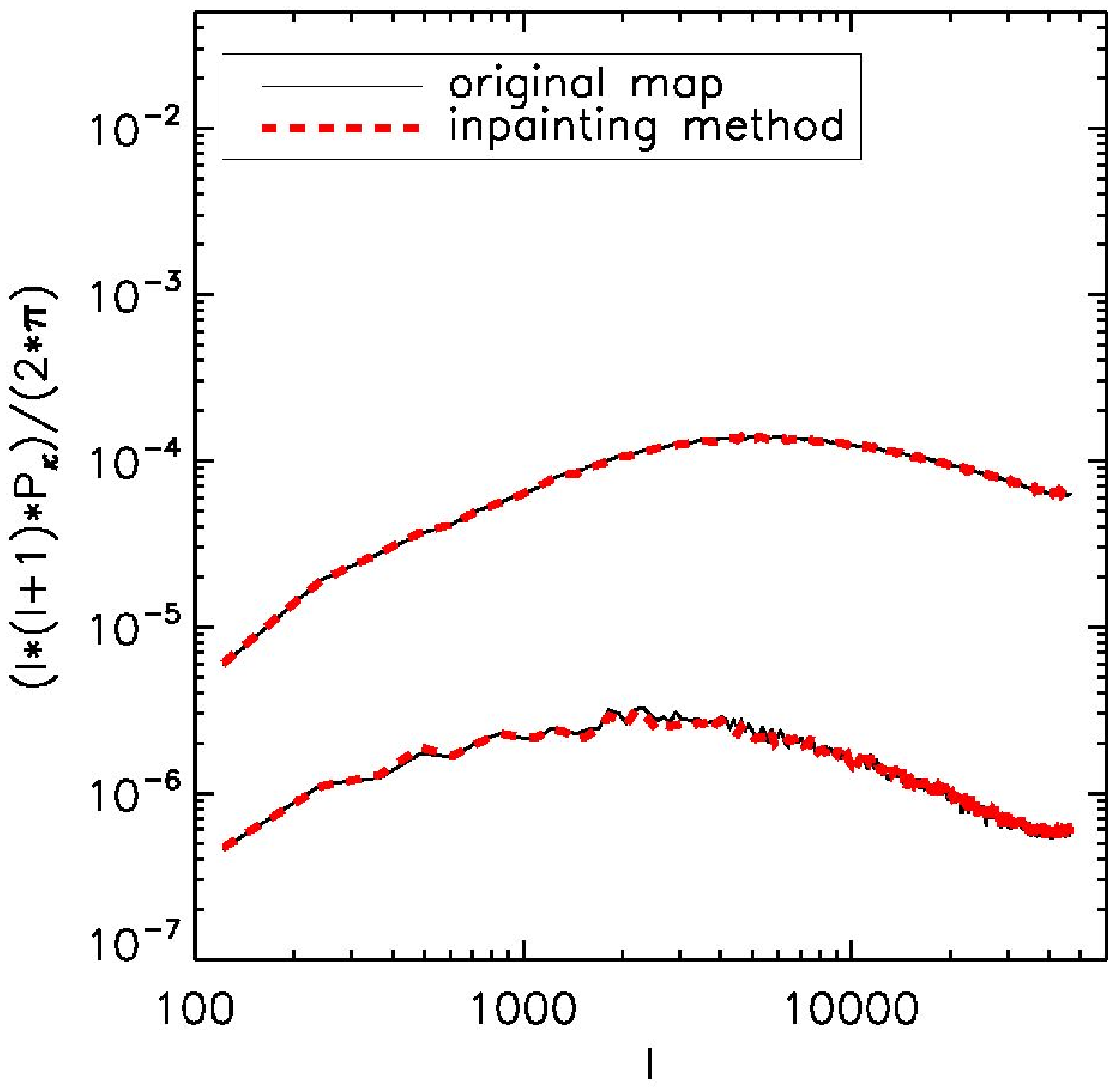}
\hspace{0.7cm}
\includegraphics[width=8.cm, height=5.cm]{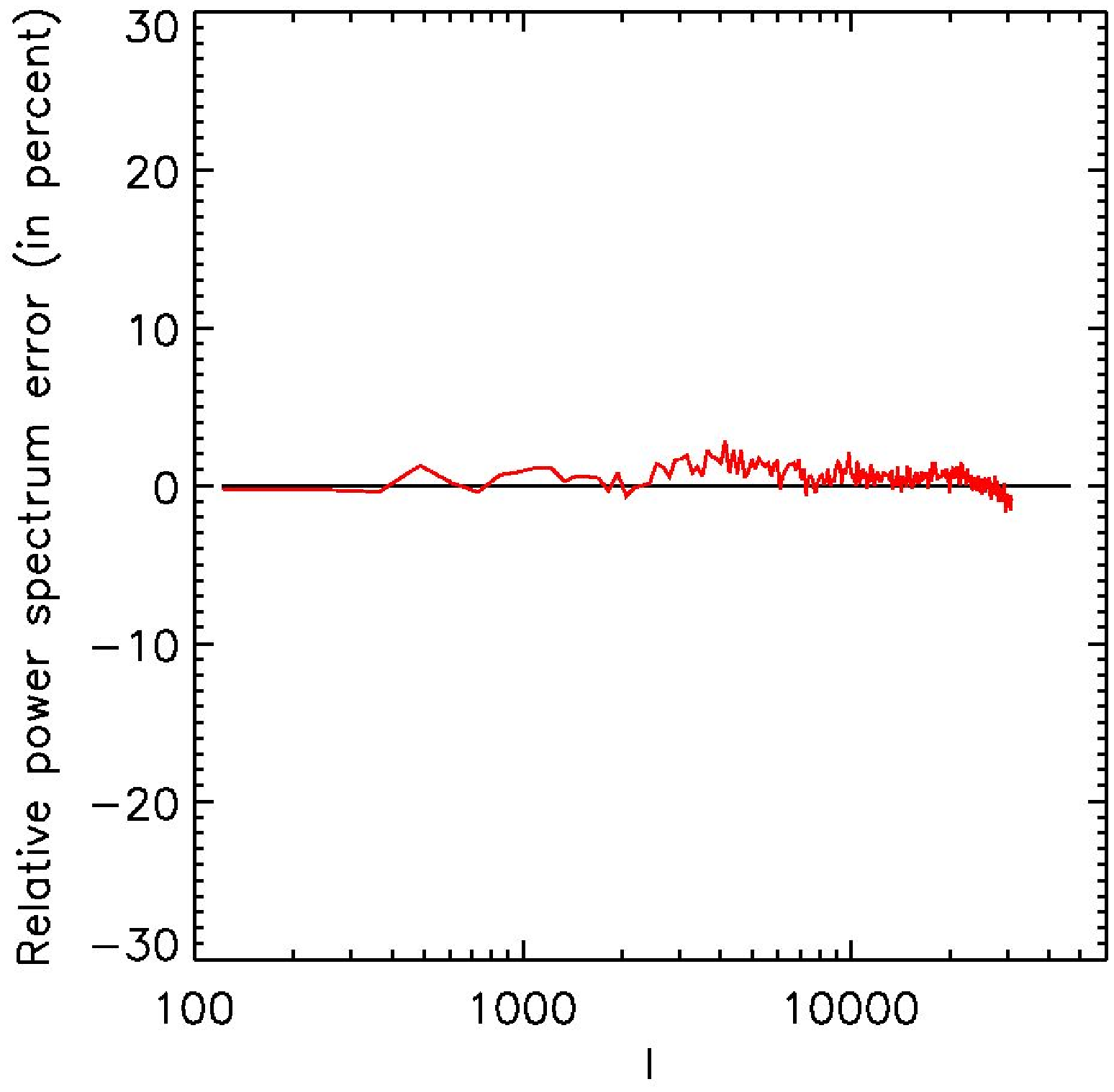}
}}}
\caption{Power spectrum recovery from convergence maps for CFHTLS mask:  left, the two upper curves (almost superposed) correspond to the mean power spectrum
computed from i) the complete simulated weak lensing mass maps (black - continuous line)  and ii) the inpainted masked maps (red - dashed line), and
the two lower curves are the  empirical standard deviation for the complete maps (black - continuous line) and the inpainted masked maps (red - dashed line).
Right, relative power spectrum  error, i.e. the normalized difference between two upper curves of the left pannel.}

\label{power7}
\vbox{
\centerline{
\hbox{
\includegraphics[width=8.cm, height=5.cm]{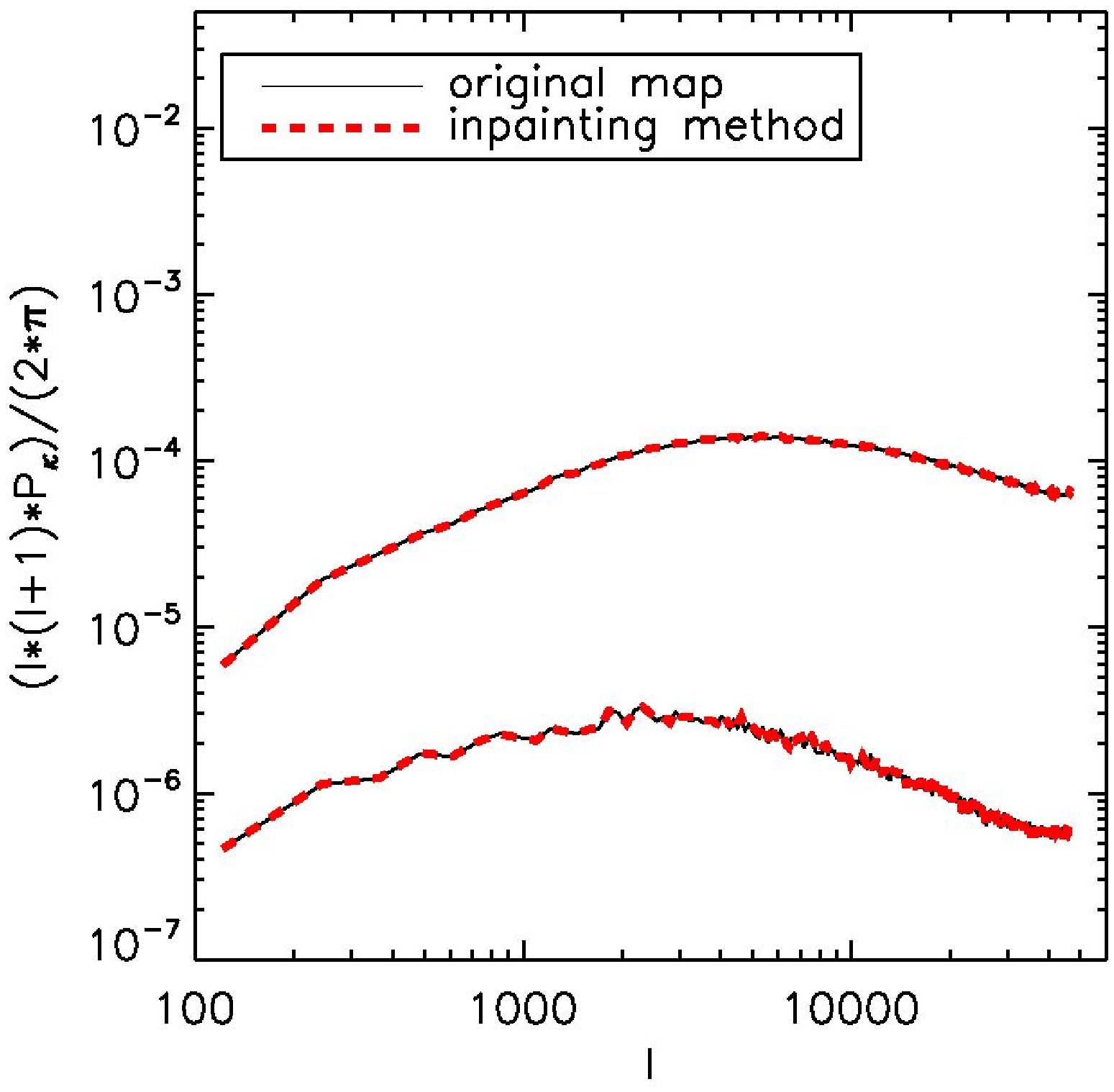}
\hspace{0.7cm}
\includegraphics[width=8.cm, height=5.cm]{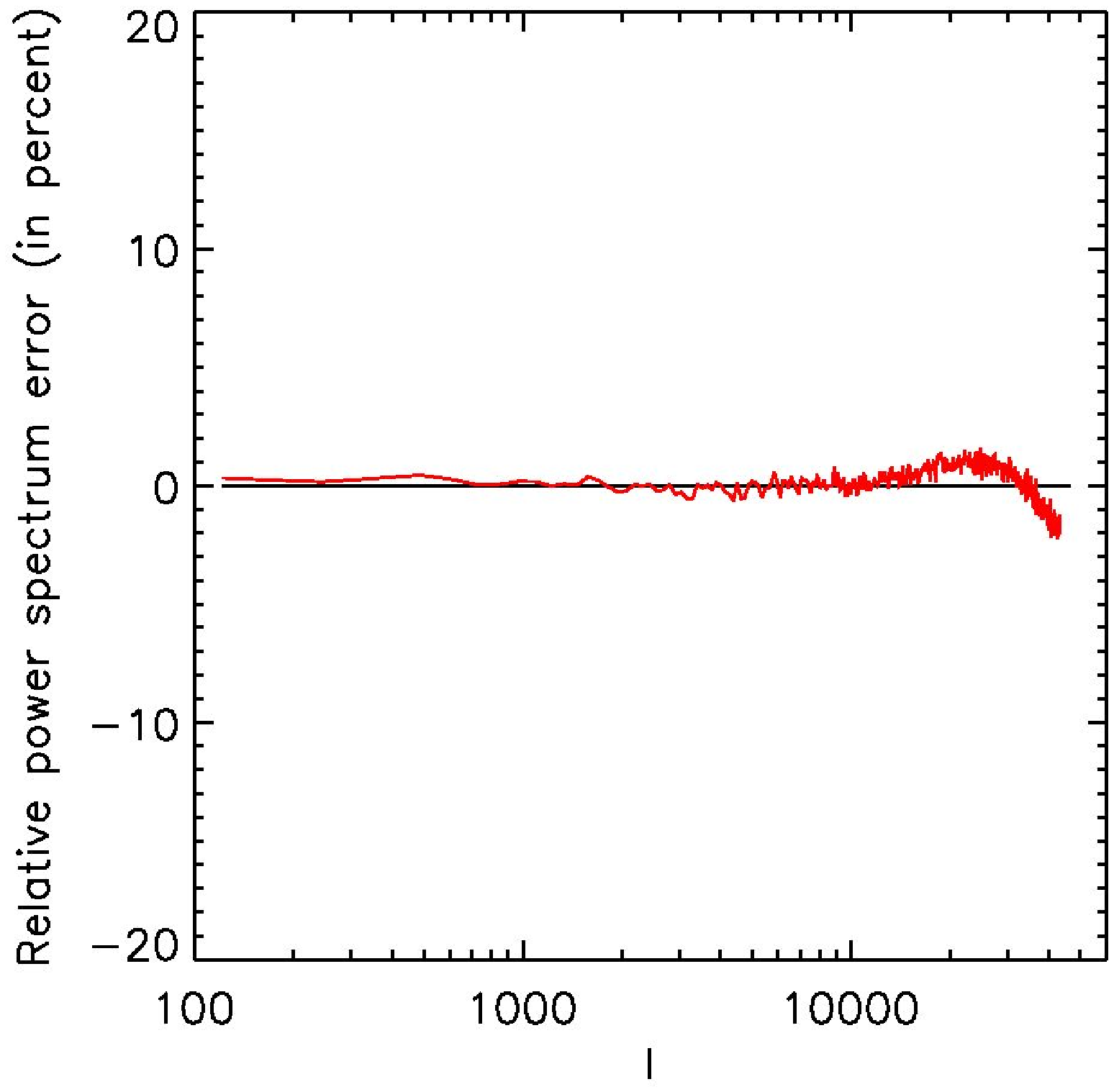}
}}
}
\caption{Power spectrum recovery from convergence maps for the Subaru mask.}
\label{power7b}
\end{figure*}

This experiment was conducted over 100 simulated incomplete mass maps.  For these maps, for both the complete maps (i.e. no gaps)
and the inpainted masked maps,  we have computed the  mean power spectrum: 
\begin{eqnarray}
<P_{\kappa}>=\frac{1}{N_m} \sum_m P_{\kappa^m}.
\label{MPS}
\end{eqnarray}
where $m$ is the number of simulations, the empirical standard deviation also called (the square root of) sample variance:
\begin{eqnarray}
\sigma_{P_{\kappa}}=\sqrt{\frac{1}{N_m} \sum_m (P_{\kappa^m} - <P_{\kappa}>)^2}.
\label{STDPS}
\end{eqnarray}
and the relative power spectrum error $E^R_{P_{\kappa}}$:
\begin{eqnarray}
E^R_{P_{\kappa}}=\frac{1}{N_m} \sum_m \left( \frac{P_{\kappa^m} - P_{\tilde{\kappa}^m} }{P_{\kappa^m} }\right).
\label{RPSEb}
\end{eqnarray}

This experiment was done for the  two kinds of mask (CFHTLS and Subaru).
 
Fig. \ref{power7} shows the results for the  CFHTLS mask.
The left panel shows four curves: the two upper curves (almost superimposed) correspond to the mean power spectrum
computed from i) the complete simulated weak lensing mass maps (black - continuous line)  and ii) the inpainted masked maps (red - dashed line).
The two lower curves are the  empirical standard deviation for the complete maps (black - continuous line) and the inpainted masked maps (red - dashed line).
Fig. \ref{power7} right shows the relative power spectrum error, i.e. the normalized difference between the two upper curves of the left pannel. 
Fig. \ref{power7b} shows the same plots  for the  Subaru mask.

We can see that  the maximum discrepancy is obtained for $l > 10000$ with Subaru mask  where the relative power spectrum error is about $3\%$.

\subsubsection*{Computing time:  2PCF  versus  the  inpainting-power spectrum method}

The two point-correlation function estimator is based on the notion of pair counting (see $\S$ \ref{two-point}). It is not biased by missing data, but
its computational time is long. On our simulated field that covers a region of $1.975^\circ \times 1.975^\circ$, 8 hours are needed to process the two point correlation function in all the field with bins having the pixel size on a 2.5 GHz processor PC-linux using C++. The proposed algorithm aims at lowering the impact of masked stars in the field while keeping fast calculation. The time to compute the power spectrum including the MCA-inpainting method in the same field still using the same 2.5 GHz processor PC-linux and C++ language is only 4 minutes. This is 120 times faster than the two-point correlation function.  It only requires $O(N \log N)$ operations compared to the two-point correlation estimation that requires $O(N^2)$ operations.

\section{Reconstruction of Weak Lensing mass maps from incomplete shear maps}
\label{sect_inp_kappa}

In previous sections, we have investigated the impact of masking out the convergence field which is a good first approximation. However, in real data, galaxy images are used to measure the shear field. This shear field can then be converted into a convergence field (see eq. \ref{eq:gamma_psi} and eq. \ref{eq:kappa_psi}). The mask for saturated stars is therefore applied to the shear field (i.e. the initial data), and we want to reconstruct the dark matter mass map $\kappa$ from the incomplete shear field $\gamma$.

\subsection{The inverse problem}
\label{wl_inversion}
In weak lensing surveys, the shear $\gamma_i({\mathbf \theta})$ with
$i=1,2$ is derived from the shapes of galaxies at positions ${\mathbf
\theta}$ in the image. The shear field $\gamma_i({\mathbf \theta})$
can be written in terms of the lensing potential $\psi({\mathbf
\theta})$ as (see eg. \cite{wlens:bartelmann01}):
\begin{eqnarray}
\label{eq:gamma_psi} 
\gamma_1 & = & \frac{1}{2}\left( \partial_1^2 -
\partial_2^2 \right) \psi \nonumber \\ \gamma_2 & = & \partial_1
\partial_2 \psi,
\end{eqnarray}
where the partial derivatives $\partial_i$ are with respect to
$\theta_i$.  

The projected mass distribution is given by the effective convergence $\kappa$ that integrates
the weak lensing effect along the path taken by the light. This effective convergence 
can be written using the Born approximation of small scattering as (see eg. \cite{wlens:bartelmann01}):
\begin{equation}
\label{eq:kappa_convergence}
\kappa_e (\vec{\theta}) =  \frac{3 H_0^2 \Omega_m}{2 c^2} \int_0^w \frac{f_k(w') f_k(w-w')}{f_k(w)} \frac{\delta(f_k(w') \vec{\theta}, w')}{a(w')} dw',
\end{equation}
where $f_k(w)$ is the angular diameter distance to the co-moving radius $w$, $H_0$ is the Hubble constant, $\Omega_m$ is the density of matter, $c$ is the speed of light and $a$ the expansion scale parameter, $\delta$ is the Dirac distribution.

The projected mass distribution $\kappa({\mathbf \theta})$ can also be
expressed in terms of the lensing potential $\psi$ as :
\begin{equation}
\label{eq:kappa_psi}
\kappa =  \frac{1}{2}\left(\partial_1^2 + \partial_2^2 \right) \psi.
\end{equation}
 
The weak lensing mass inversion problem consists of reconstructing the
projected (normalized) mass distribution $\kappa({\mathbf \theta})$
from the incomplete measured shear field $\gamma_i({\mathbf \theta})$ by
inverting equations \ref{eq:gamma_psi} and \ref{eq:kappa_psi}.  This is an ill posed problem that 
need to be regularized. 

\subsection{The sparse solution}
By taking the Fourier transform of equations~\ref{eq:gamma_psi} and \ref{eq:kappa_psi}, we have
\begin{equation}
\hat{\gamma_i} = \hat{P_i} \hat{\kappa},~~~i=1,2,
\label{eq:gamma}
\end{equation}
where the hat symbol denotes Fourier transforms. We define
$k^2 \equiv k_1^2 + k_2^2$ and
\begin{eqnarray}
\hat{P_1}(\mathbf k) & = & \frac{k_1^2 - k_2^2}{k^2} \nonumber \\
\hat{P_2}(\mathbf k) & = & \frac{2 k_1 k_2}{k^2},
\label{eq:p_inversion}
\end{eqnarray}
with $\hat{P_1}(k_1,k_2) \equiv 0$ when $k_1^2 = k_2^2$, and
$\hat{P_2}(k_1,k_2) \equiv 0$ when $k_1 = 0$ or $k_2 = 0$. 

 We can easily derive an estimation of the mass map by inverse filtering, 
 the least-squares estimator $\tilde{\kappa}$ of the convergence $\kappa$ is e.g. \citep{starck:sta05}:
 \begin{eqnarray}
 \tilde{\kappa} & = & P_1* \gamma_{1}+ P_2*\gamma_{2}.
 \label{eq_reckE}
\end{eqnarray}

We have ${\gamma_i} = {P_i} * {\kappa}$,  where $*$ denotes convolution. When the data are not complete, we have:
\begin{equation}
\gamma_i = M (P_i * \kappa),~~~i=1,2,
\label{eq:gamma_gaps}
\end{equation}

To treat masks applied to shear field, the dictionary $\Phi$ is unchanged because the DCT remains the best representation for the data, but we now want to minimize: 
\begin{equation}
\min_{\kappa}  \| \Phi^T \kappa  \|_0 \quad   \text{subject to}  \quad   \sum_i \parallel \gamma_i - M (P_i * \kappa)   \parallel^2 \le \sigma.
\end{equation}

Thus, similarly to eq.~\ref{minimisation}, we can obtain the  mass map $\kappa$  from shear maps $ \gamma_i$
using the following iterative algorithm.

\subsection{Algorithm}

\begin{center}
\begin{tabular}{|c|} \hline
\begin{minipage}[b]{8.cm}
\vspace{0.1in}
\small{
\textsf{1. Set the maximum number of iterations $I_{max}$, the solution $\kappa^{0}=0$, the residual
$R^{0}=P_1*\gamma_1^{obs} + P_2*\gamma_2^{obs}$ see eq. \ref{eq_reckE} and $\gamma^{obs} = (\gamma_1^{obs}, \gamma_2^{obs})$, the maximum threshold $\lambda_{max} = \mathrm{\max}(|\alpha = \phi^T Y|$), the minimum threshold
$\lambda_{min}=0$.}

\textsf{2. Set $n$ to $0$, $\lambda_n = \lambda_{max}$. Iterate:}

\hspace{0.2cm} \textsf{3. $U = \kappa^n + M R^n(\gamma^{obs})$  and}

\hspace{0.2cm} \textsf{$R^n(\gamma^{obs}) =  P_1*(\gamma_1^{obs} - P_1*\kappa^n) + P_2*(\gamma_2^{obs} - P_2*\kappa^n) $}

\hspace{0.2cm} \textsf{4. Forward transform of $U$: $\alpha =\Phi^T U$.}

\hspace{0.2cm} \textsf{5. Determination of the threshold level }

\hspace{0.2cm} \textsf{$\lambda_n = F(n, \lambda_{max}, \lambda_{min})$.}

\hspace{0.2cm} \textsf{6. Hard-threshold the coefficient $\alpha$ using $\lambda_n$ : $\tilde{\alpha}=S_{\lambda_n}\{\alpha\}$.}

\hspace{0.2cm} \textsf{7. Reconstruct $\tilde U$ from the thresholded $\alpha$ and $\kappa^{n+1}=\Phi \tilde{\alpha}$.}

\hspace{0.2cm} \textsf{8. $n=n+1$ and if $n <  I_{max}$, return to Step 3.}
}
\vspace{0.05in}
\end{minipage}
\\\hline
\end{tabular}
\\ \vspace{0.1in}
\end{center}

$\Phi^T$ is the DCT operator. The residual $R_n$ is estimated from shear maps $\gamma_1^{obs}$ and $\gamma_2^{obs}$. Consequently, we need to use two FFTs at each iteration $n$ to compute the mass map $\kappa$ from the shear fields (eq. \ref{eq_reckE}) and the shear fields from the mass map  $\kappa$ (eq. \ref{eq:gamma}). $F$ follows the same decreasing law described in \S \ref{algorithm1}

\subsection{Experiments}
\label{inp_shear}
One of the central goals of weak lensing analysis is the measurement of cosmological parameters. 
To constrain the large-scale structures, the power spectrum $P_{\kappa}$ of the convergence $\kappa$ and thus its two-point correlation function $C_{\kappa \kappa}$, contains all the information about the primordial fluctuations. To characterize the non-gaussianity at small scales due to the growth of structures, higher-order statistics like three-point statistics have to be used. To lower the impact of the mask, we have applied the previous algorithm with 100 iterations and using always one single representation, the DCT. Then we have conducted several experiments to estimate the quality of the method.

\subsubsection*{Dark matter mass map reconstruction from incomplete shear maps}

\begin{figure*}
\vbox{
\centerline{
\hbox{
\includegraphics[width=6.5cm, height=6.5cm]{./EPS/rwl_map256_3_m5.eps}
\hspace{0.2cm}
\includegraphics[width=6.5cm, height=6.5cm]{./EPS/rwl_map256_1_m2.eps}
}}
\vspace{0.2cm}
\centerline{
\hbox{
\includegraphics[width=6.5cm, height=6.5cm]{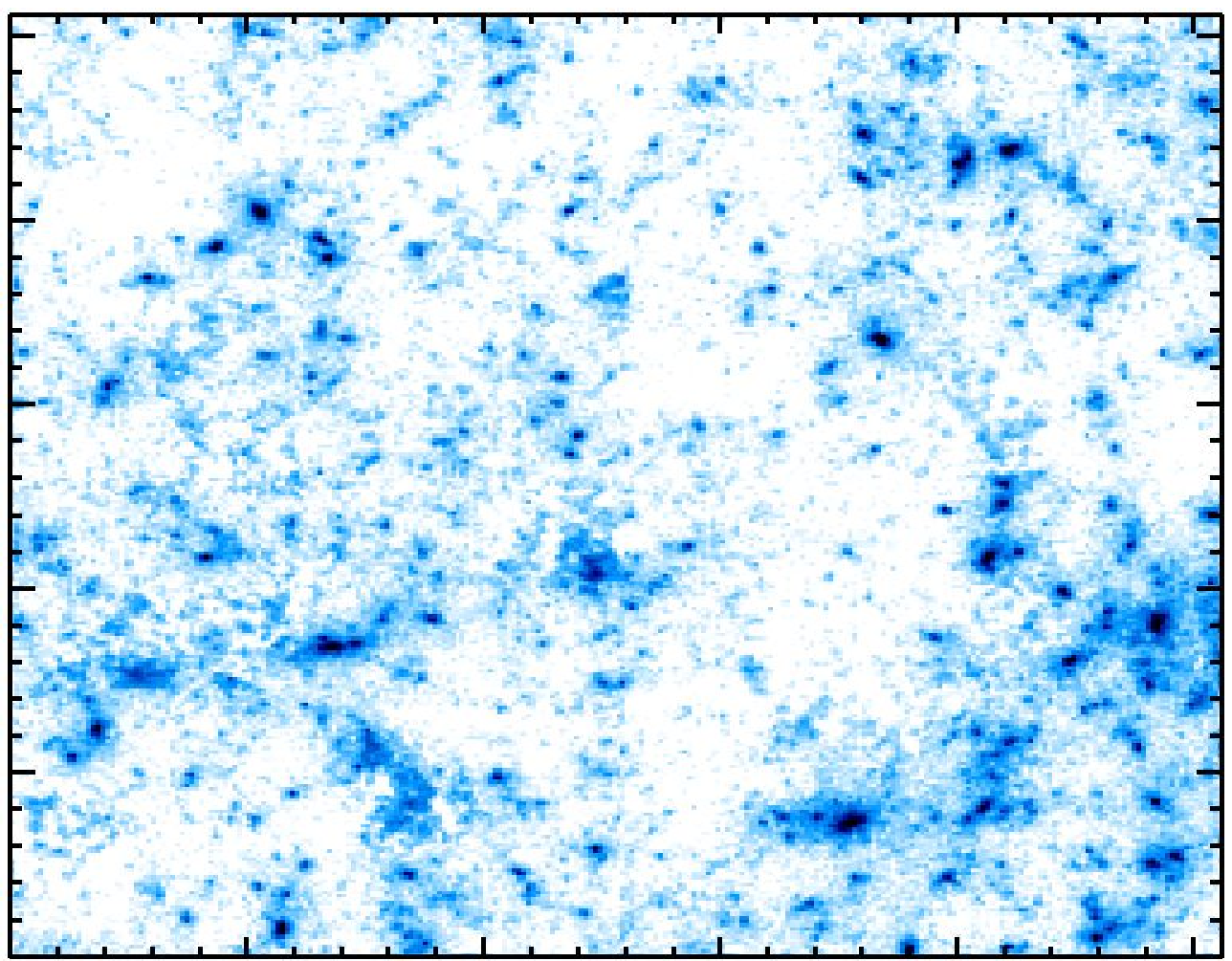}
\hspace{0.2cm}
\includegraphics[width=6.5cm, height=6.5cm]{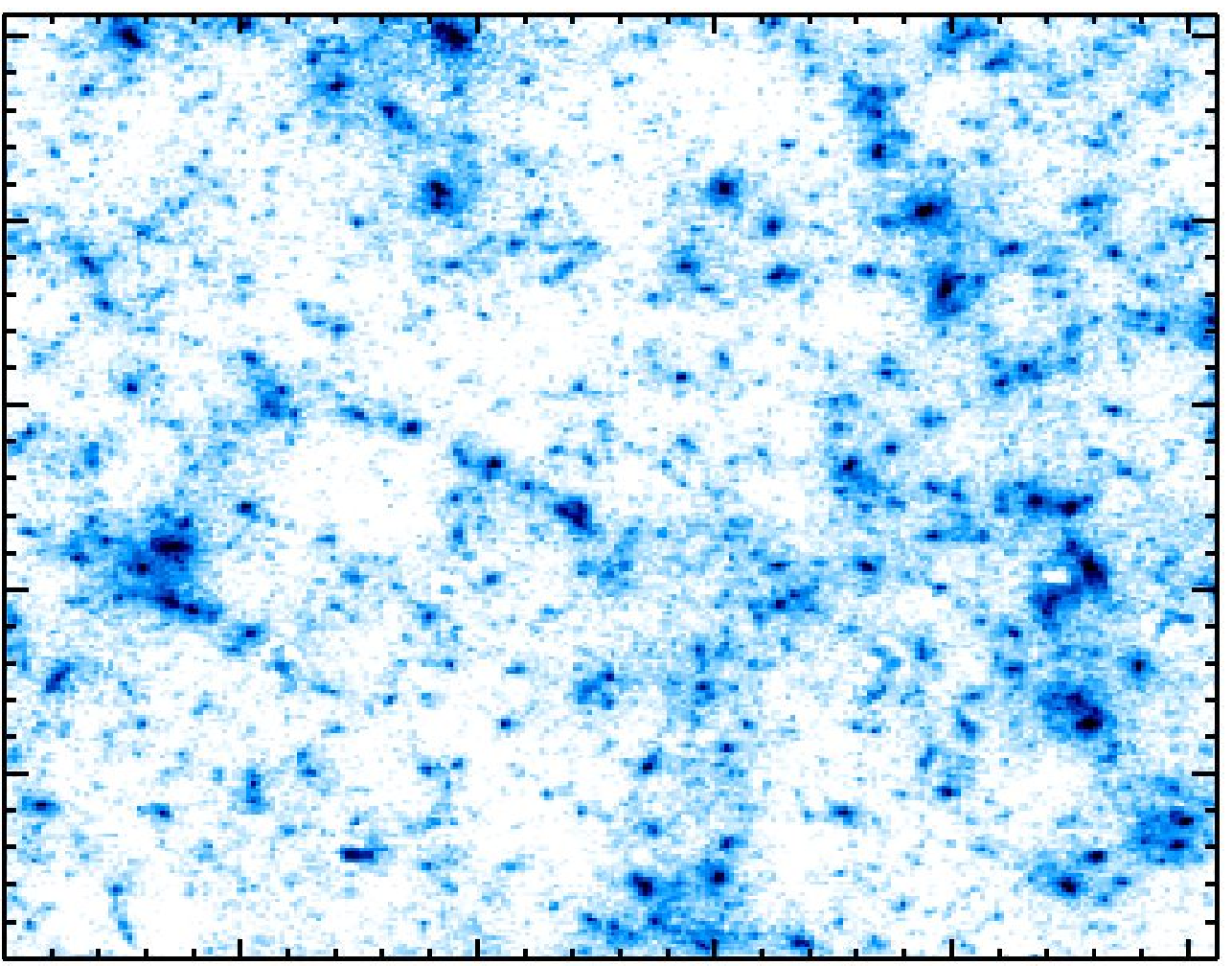}
}}
}
\caption{Upper panels, simulated weak lensing mass maps with the mask pattern of CFHTLS data on D1 field (left) and with the mask pattern of Subaru data in the same field (right) whose masks for saturated stars are applied to the shear field. Lower panels, inpainted mass maps. The region shown is $1^\circ$ x $1^\circ$.}
\label{visual}
\end{figure*}

As in the previous section, we have applied the MCA-inversion method on two different simulated weak lensing mass maps whose shear field have been masked by the two typical mask patterns.  
Fig. \ref{visual}, top panels, shows the simulated mass map masked by the CFHTLS mask (on the left) and by the Subaru mask (on the right). The results applying the MCA-inpainting method described above is shown in the bottom panels. The gaps are again undistinguishable by eye in both cases.

\subsubsection*{Power spectrum estimation of the convergence $\kappa$ from incomplete shear maps}

\begin{figure*}
\vbox{
\centerline{
\hbox{
\includegraphics[width=8.cm, height=5.cm]{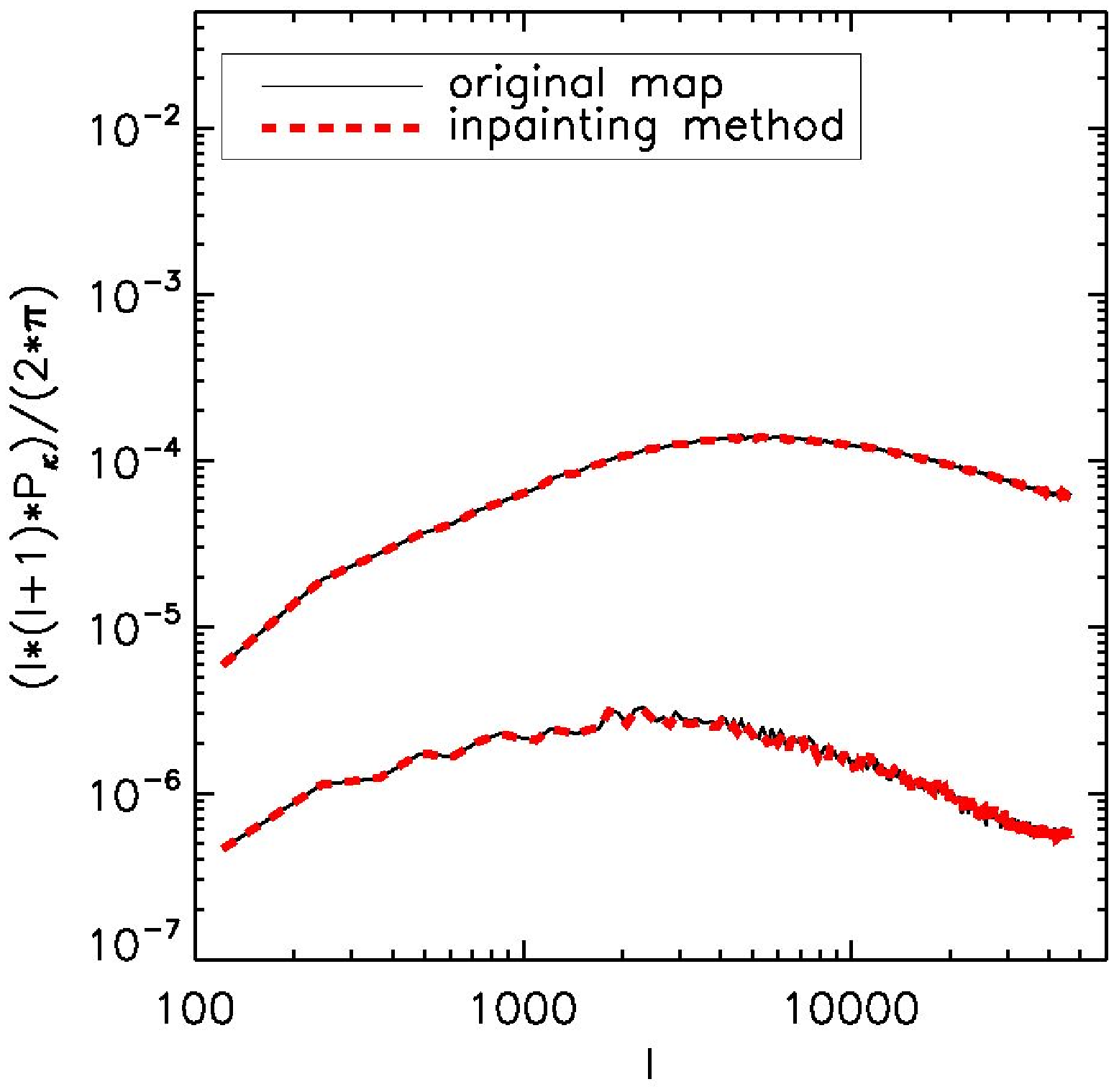}
\hspace{0.7cm}
\includegraphics[width=8.cm, height=5.cm]{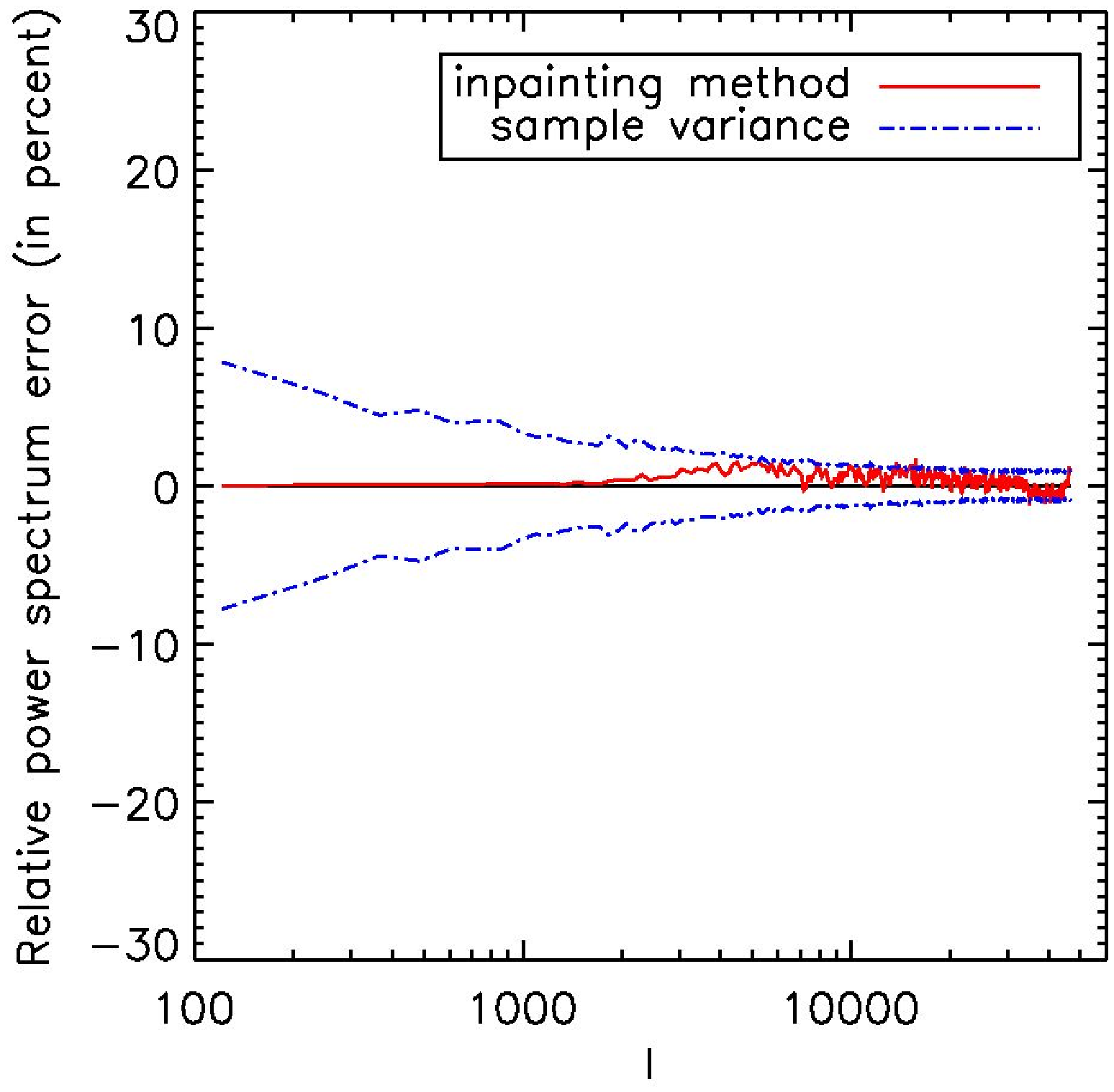}
}}}
\caption{Power spectrum recovery from shear maps for CFHTLS mask:  left, the two upper curves (almost superposed) correspond to the mean power spectrum computed from i) the complete simulated weak lensing mass maps (black - continuous line)  and ii) the inpainted reconstructed maps from masked shear maps (red - dashed line), and the two lower curves are the  empirical standard deviation for the complete maps (black - continuous line) and the inpainted reconstructed maps from masked shear maps (red - dashed line). Right, relative power spectrum  error, i.e. the normalized difference between two upper curves of the left pannel. The blue - dashed line represents the empirical standard deviation (cosmic variance) estimated from the complete mass maps.}
\label{power8}

\vbox{
\centerline{
\hbox{
\includegraphics[width=8.cm, height=5.cm]{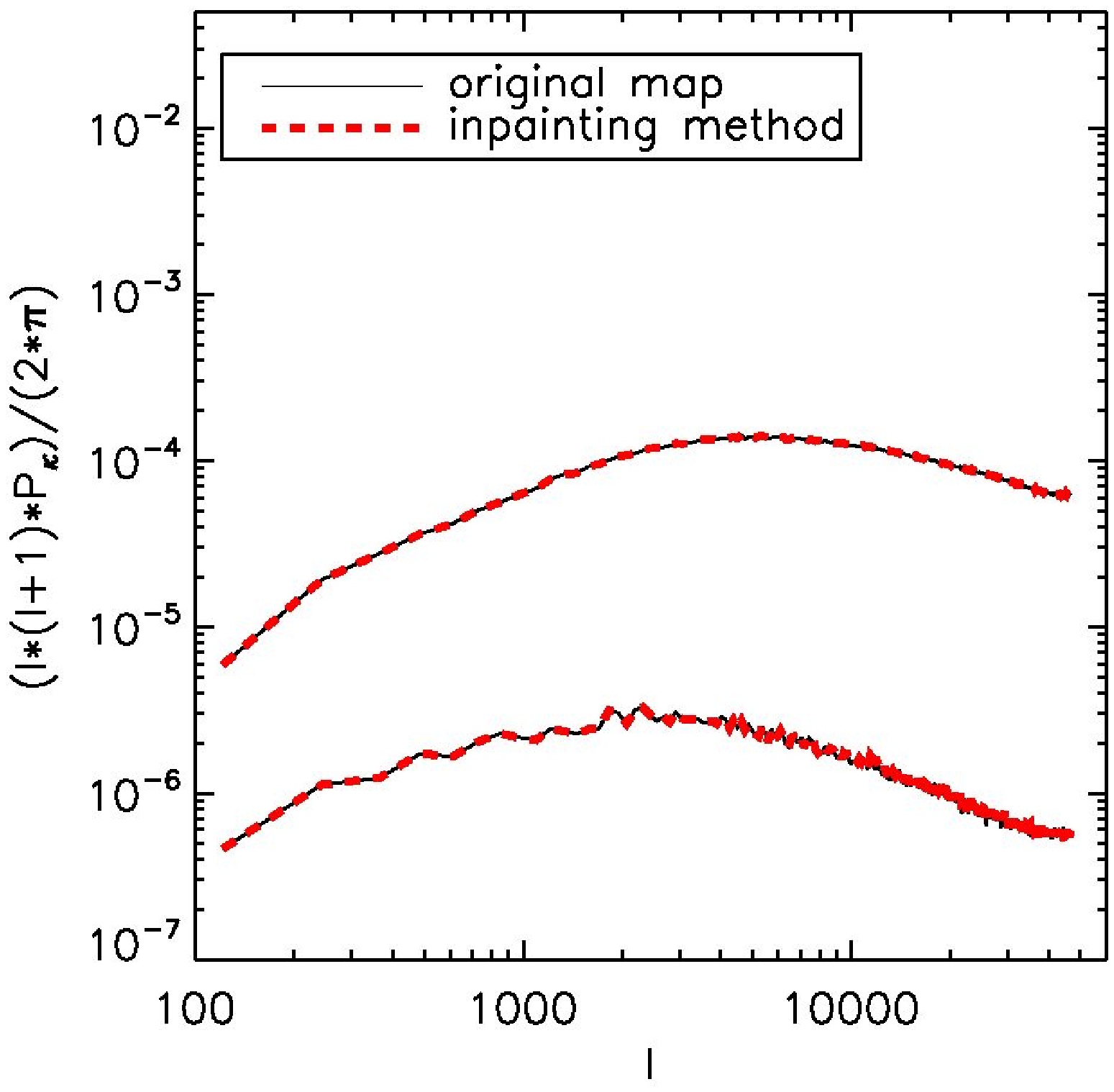}
\hspace{0.7cm}
\includegraphics[width=8.cm, height=5.cm]{./EPS/diff_powspec100_g_m5.eps}
}}
}
\caption{Power spectrum recovery from shear maps for the Subaru mask.}
\label{power8b}
\end{figure*}

As in the previous section, this experiment was done over 100 incomplete shear maps for the two kinds of mask (CFHTLS and Subaru).
 
Fig. \ref{power8} shows the results for the  CFHTLS mask.
The left panel shows four curves: the two upper curves (almost superposed) correspond to the mean power spectrum
computed from i) the complete simulated weak lensing mass maps (black - continuous line)  and ii) the inpainted reconstructed maps from masked shear maps (red - dashed line). The two lower curves are the empirical standard deviation estimated from the complete maps (black - continuous line) and the inpainted reconstructed maps from masked shear maps (red - dashed line).
Fig. \ref{power8} right shows the relative power spectrum error, i.e. the normalized difference between the two upper curves of the left pannel. And the blue - dashed  line is the root mean square of the sample variance that comes from the finite size of the field. 
Fig. \ref{power8b} shows the same plots  for the  Subaru mask.

We can see that  the maximum discrepancy is obtained in the $l$-range of $[2000, 7000]$  with CFHTLS mask  where the relative power spectrum error is about $1\%$ while is about $0.3 \%$ with Subaru mask. The error introduced by our method is consistent with the sample variance.

\subsubsection*{Case of noisy incomplete shear maps}

\begin{figure*}
\vbox{
\centerline{
\hbox{
\includegraphics[width=8.cm, height=5.cm]{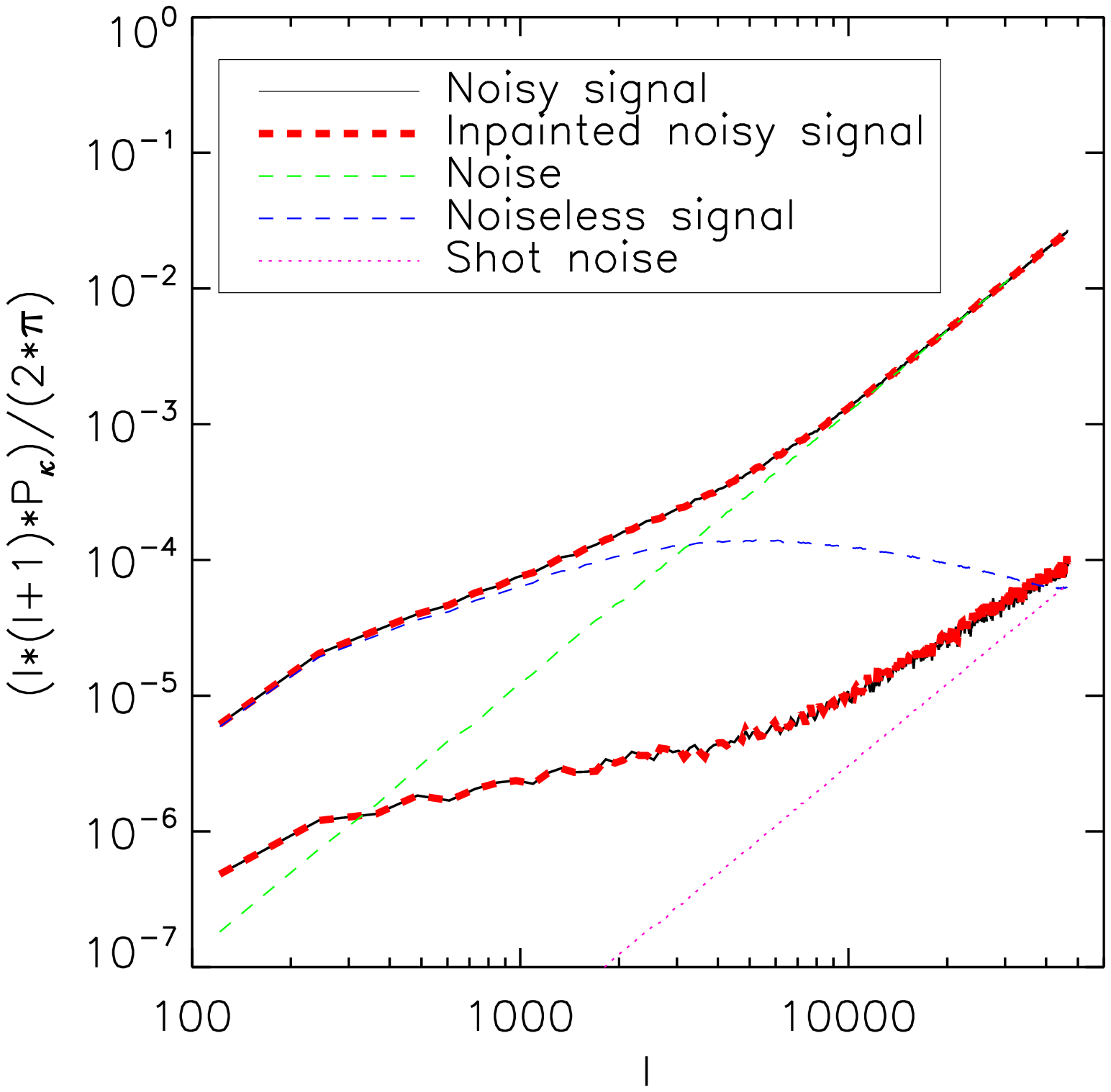}
\hspace{0.7cm}
\includegraphics[width=8.cm, height=5.cm]{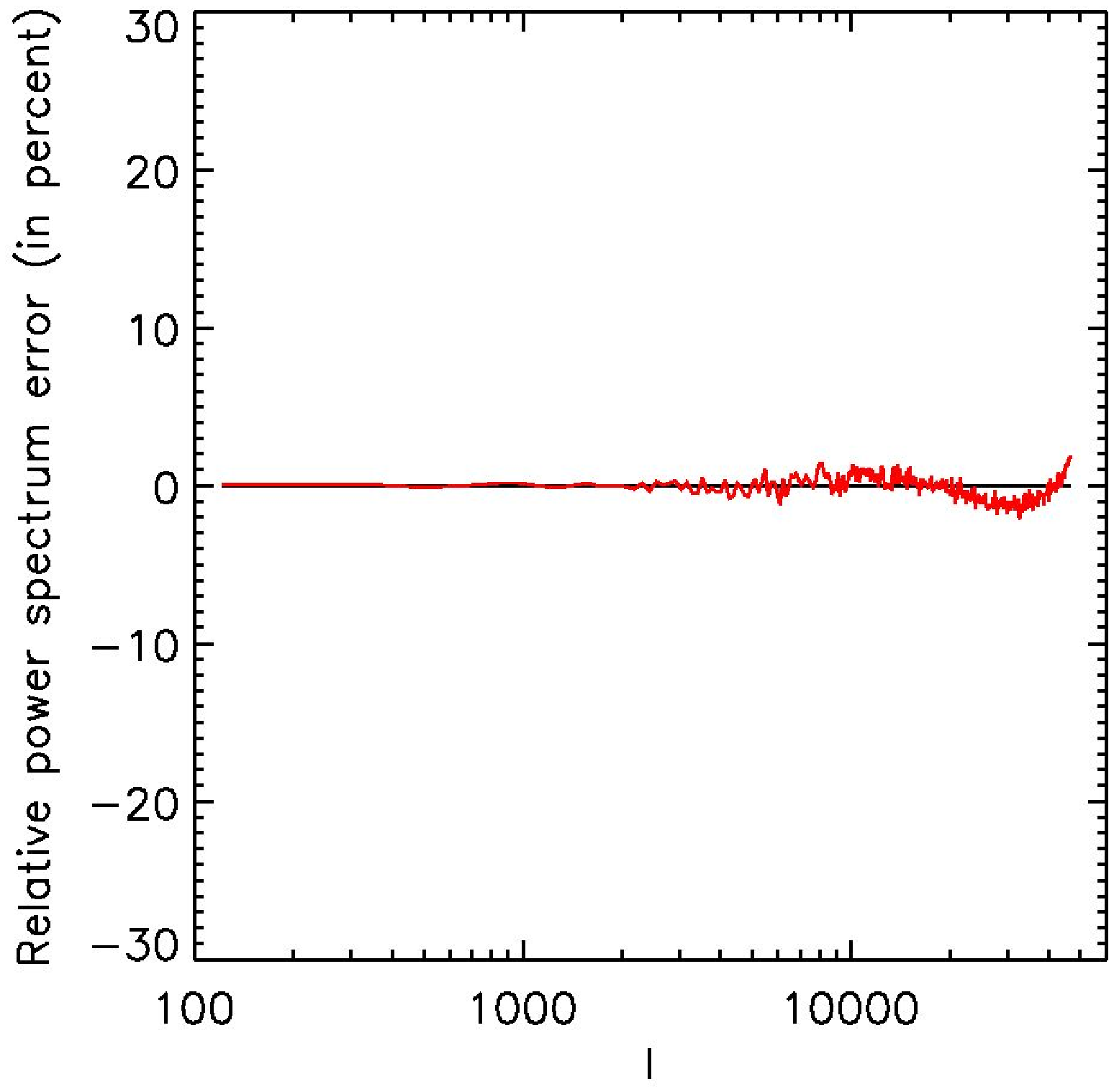}
}}}
\caption{Power spectrum recovery from noisy shear maps for CFHTLS mask:  left, the two upper curves (almost superimposed) correspond to the mean power spectrum computed from i) the complete simulated noisy mass maps (black - continuous line)  and ii) the inpainted reconstructed maps from masked noisy shear maps (red - dashed line). The mean power spectrum of the noise is plotted as a dashed green line and the mean power spectrum of the noiseless mass maps is plotted as a dashed blue line. The pink dashed line represents the simulation shot noise. The two lower curves are the  empirical standard deviation for the complete noisy mass maps (black - continuous line) and the inpainted reconstructed maps from masked noisy shear maps (red - dashed line). Right, relative power spectrum error, i.e. the normalized difference between two upper curves of the left pannel.}
\label{power9}
\vbox{
\centerline{
\hbox{
\includegraphics[width=8.cm, height=5.cm]{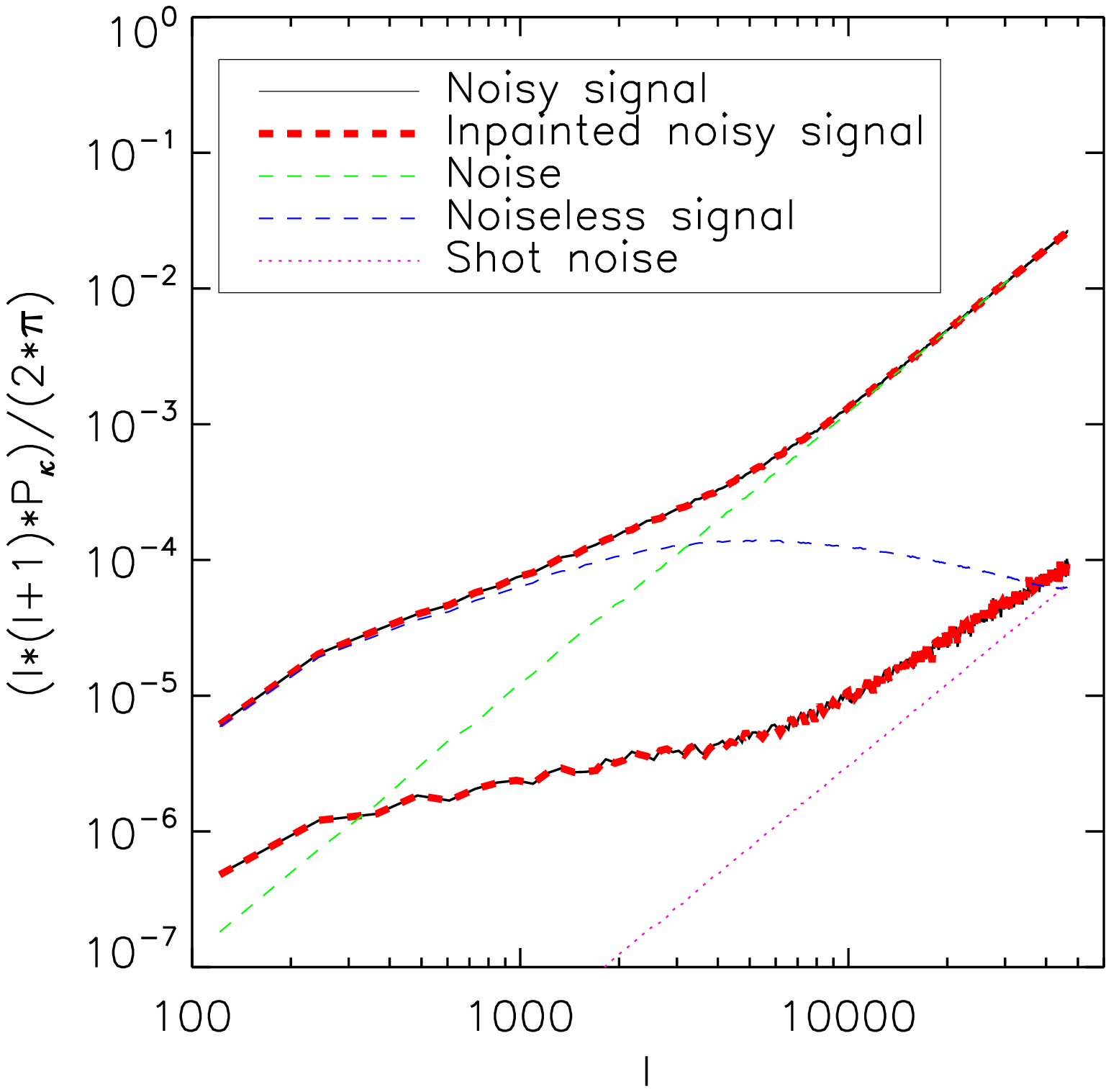}
\hspace{0.7cm}
\includegraphics[width=8.cm, height=5.cm]{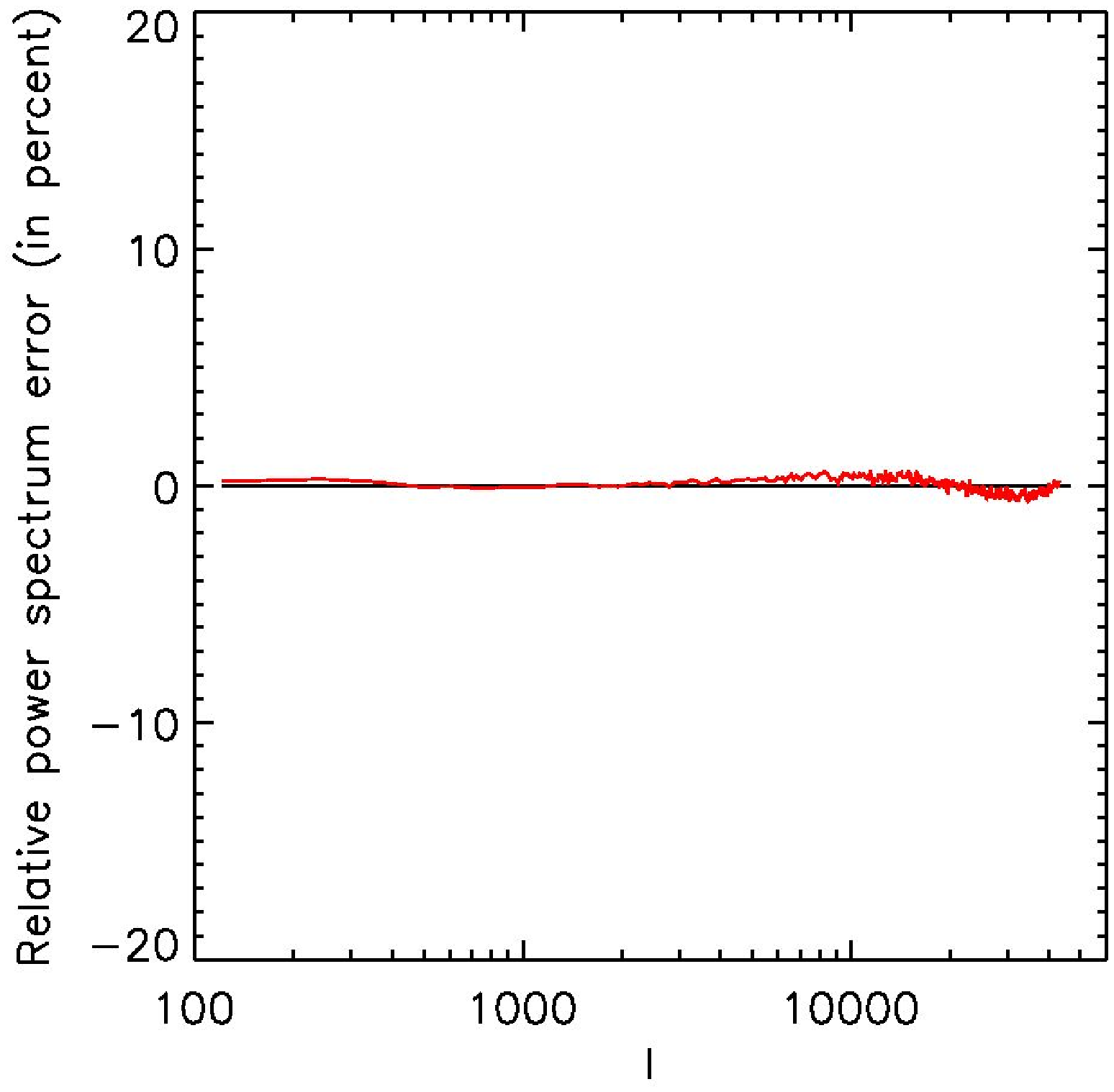}
}}
}
\caption{Power spectrum recovery from noisy shear maps for the Subaru mask.}
\label{power9b}
\end{figure*}

As discussed in section \ref{noise}, if the data contains some noise, it is straightforward to take it into account using a similar processing. The weak lensing data are noisy  because the observed shear $\gamma_i$ is obtained by averaging over a finite number of galaxies.
Another experiment has been conducted still using 100 simulated weak lensing mass maps with noise to simulate space observations (100 galaxies/arcmin$^2$ and with the shear error per galaxy given by $\sigma_{\gamma}=0.3$). The two kinds of mask (CFHTLS and Subaru) have been also tested.
 
Fig. \ref{power9} shows the results for the  CFHTLS mask.
Left panel, the two upper curves (almost superposed) correspond to the mean power spectrum
computed from i) the complete simulated noisy mass maps (black - continuous line)  and ii) the inpainted maps from masked noisy shear maps (red - dashed line). The noise power spectrum represented in green dashed line modify the shape of the noiseless weak lensing power spectrum plotted in blue dashed line. The two lower curves are the  empirical standard deviation for the complete noisy mass maps (black - continuous line) and the inpainted maps from masked noisy shear maps (red - dashed line).
Fig. \ref{power9} right shows the relative power spectrum error, i.e. the normalized difference between two upper curves of the left pannel. Fig. \ref{power9b} shows the same plots  for the  Subaru mask.

We can see that the maximum discrepancy is obtained with CFHTLS mask in the $l$-range of [25000, 40000]  where the relative power spectrum error is about $1\%$ and is only $0.5\%$ with Subaru mask. The error is not amplified by the noise.

\subsubsection*{ Equilateral bispectrum estimation of the convergence $\kappa$ from incomplete shear maps}

\begin{figure*}
\vbox{
\centerline{
\hbox{
\includegraphics[width=8.cm, height=5.cm]{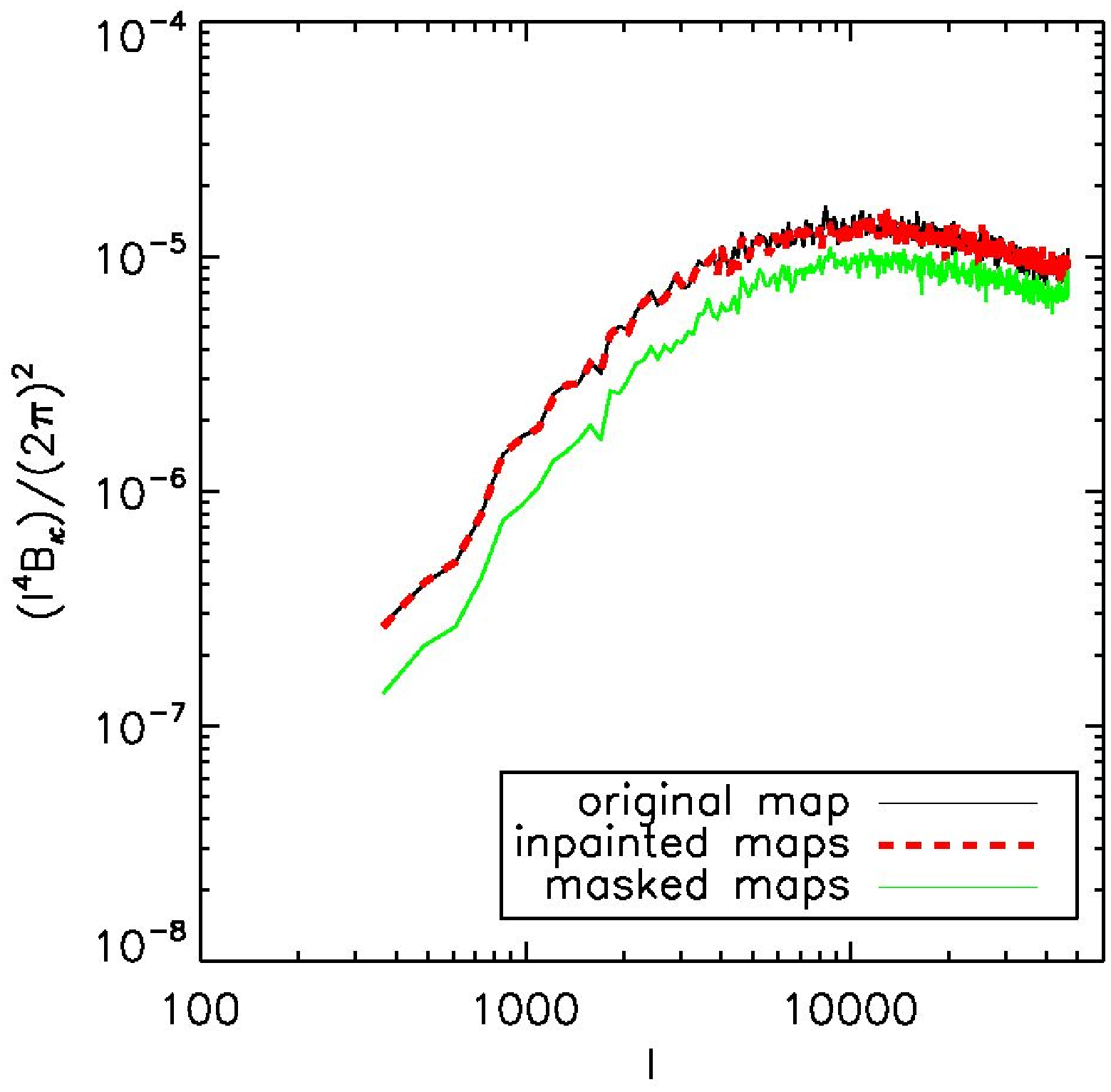}
\hspace{0.7cm}
\includegraphics[width=8.cm, height=5.cm]{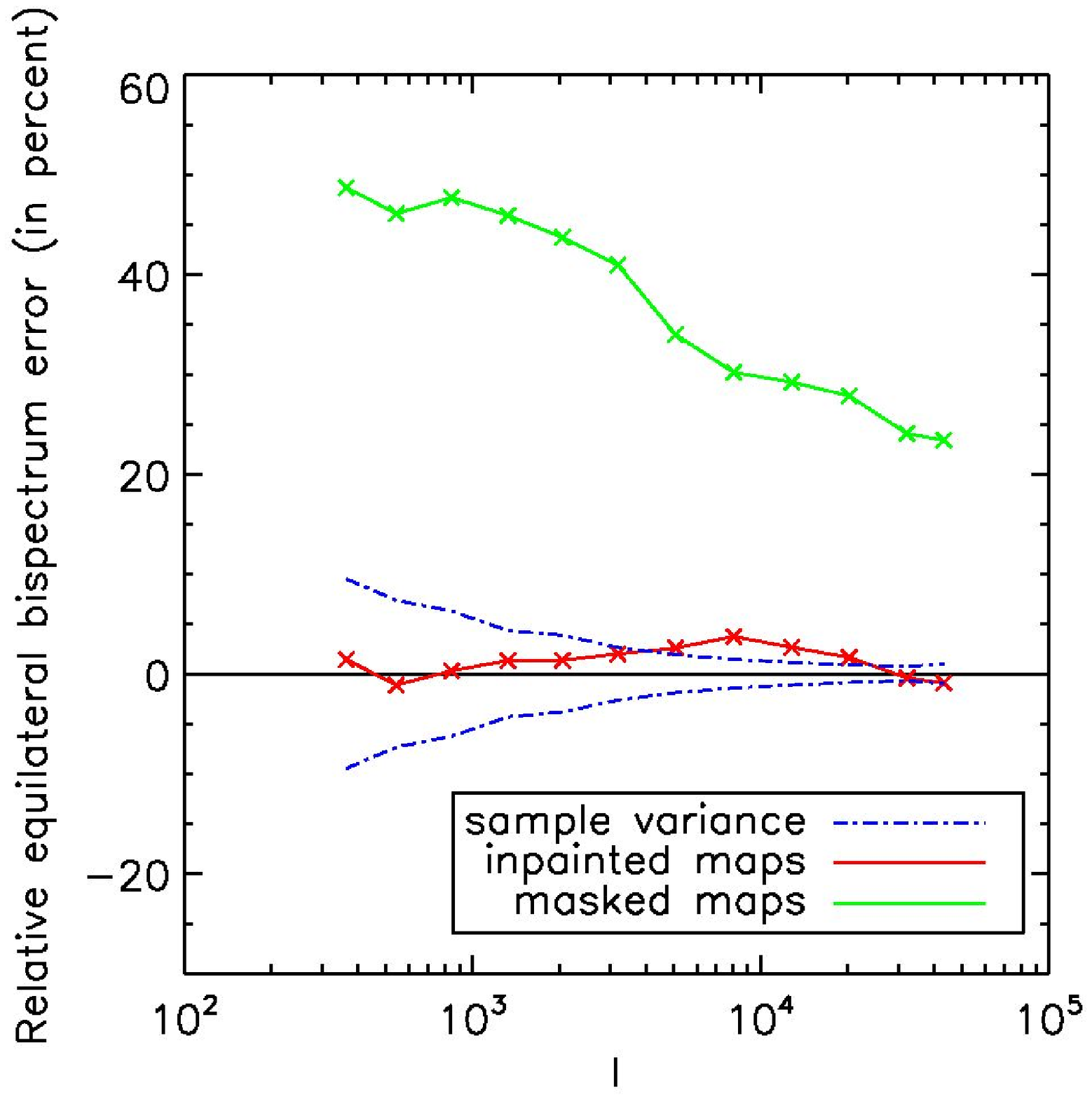}
}}}
\caption{Bispectrum recovery for CFHTLS mask:  left, the two upper curves (almost superposed) correspond to the mean equilateral bispectrum computed from i) the complete simulated weak lensing mass maps (black - continuous line)  and ii) the inpainted maps from masked shear maps (red - dashed line), and the lower curve correspond to the mean equilateral bispectrum computed from the incomplete shear maps (green - continuous line). Right, relative power spectrum error, i.e. the normalized difference between the curves of the left panel in logarithmic bins. The blue - dashed line represents the empirical standard deviation estimated from the complete mass maps, previously on black - continuous line.}

\label{power10}
\vbox{
\centerline{
\hbox{
\includegraphics[width=8.cm, height=5.cm]{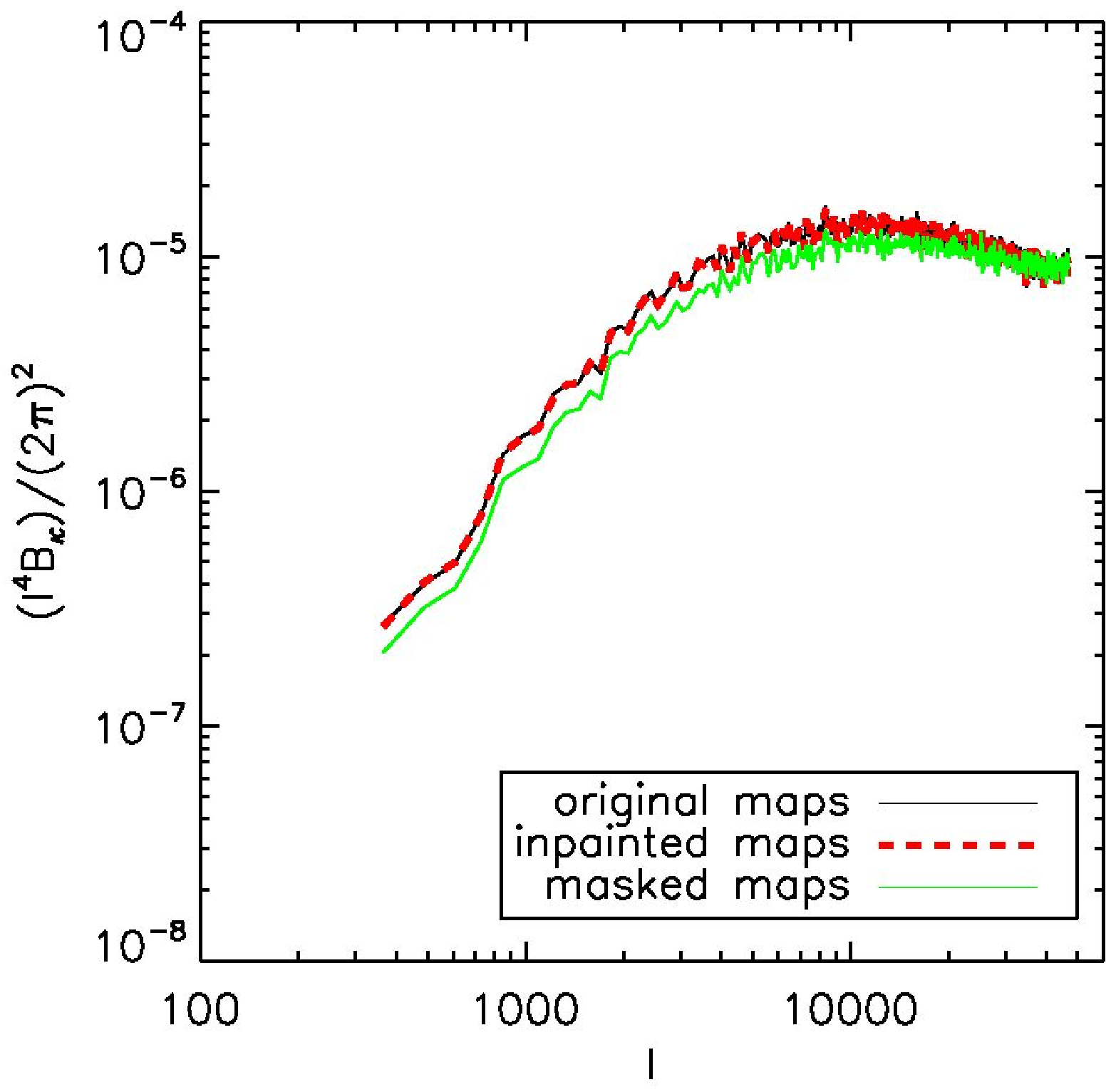}
\hspace{0.7cm}
\includegraphics[width=8.cm, height=5.cm]{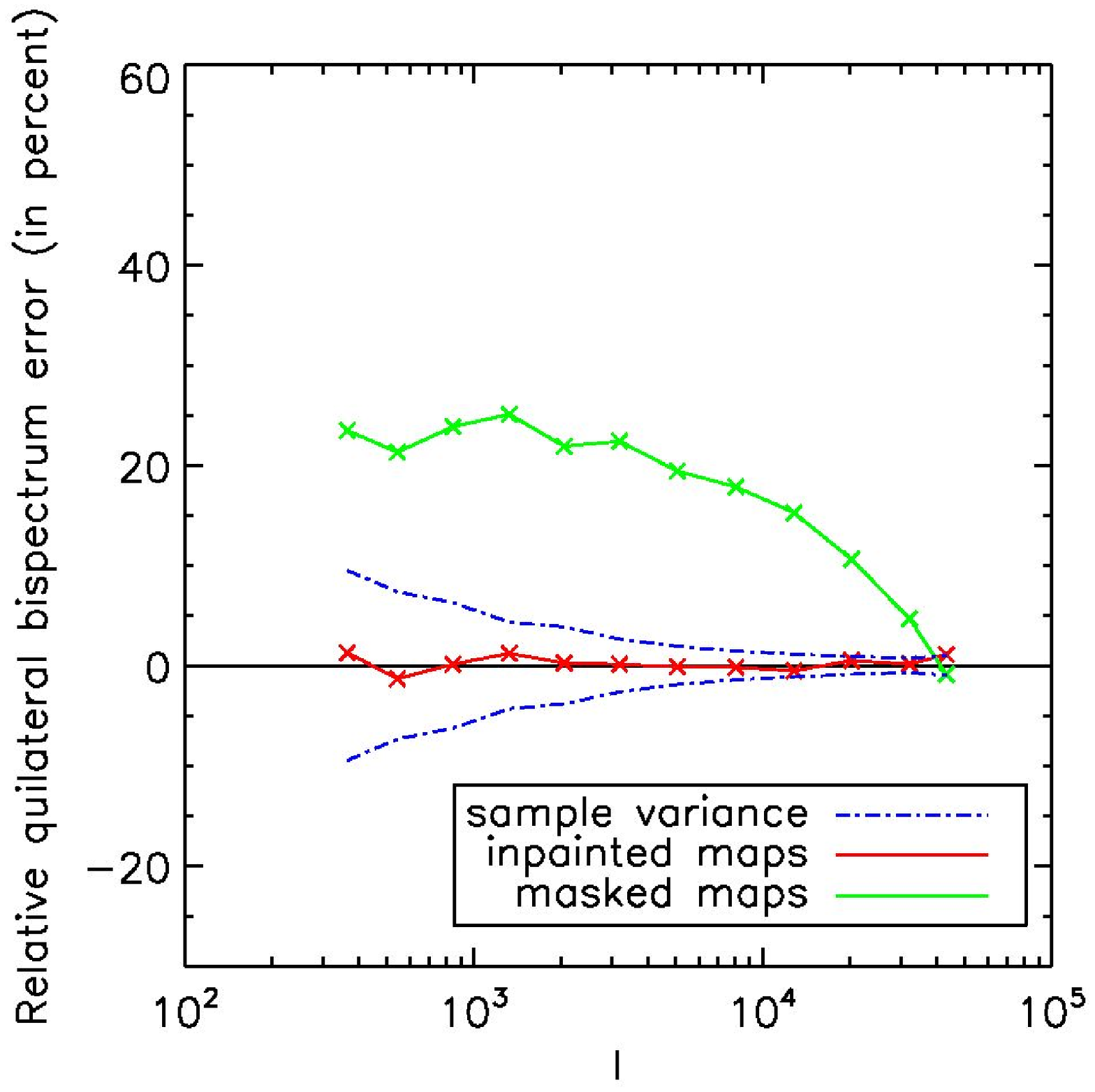}
}}
}
\caption{Bispectrum recovery for the Subaru mask.}
\label{power10b}
\end{figure*}

We consider now the equilateral bispectrum that we have introduced \S \ref{3pts}.
We defined the mean equilateral bispectrum as follows : 
\begin{eqnarray}
<B^{eq}_{\kappa}>=\frac{1}{N_m} \sum_m B^{eq}_{\kappa^m}.
\label{MPSb}
\end{eqnarray}
And the relative equilateral bispectrum error $E^R_{B^{eq}_{\kappa}}$ is given by :
\begin{eqnarray}
E^R_{B^{eq}_{\kappa}}=\frac{1}{N_m} \sum_m \left( \frac{B^{eq}_{\kappa^m} - B^{eq}_{\tilde{\kappa}^m} }{B^{eq}_{\kappa^m} }\right),
\label{RPSE}
\end{eqnarray}

The experiment was also done for the two kinds of mask (CFHTLS and Subaru).
Fig. \ref{power10} shows the results for the  CFHTLS mask.
The left panel shows three curves (whose two almost superposed) that correspond to the mean equilateral bispectrum
computed from i) the complete simulated mass maps (black - continuous line)  and ii) the inpainted maps from masked shear maps (red - dashed line) and iii) the incomplete simulated shear maps (green - continuous line).
Fig. \ref{power10} right shows the relative equilateral bispectrum error in logarithmic bins, i.e. the normalized difference between the curves of the left pannel. Fig. \ref{power10b} shows the same plots  for the  Subaru mask.

We defined the mean equilateral bispectrum as follows : 
\begin{eqnarray}
<B^{eq}_{\kappa}>=\frac{1}{N_m} \sum_m B^{eq}_{\kappa^m}.
\label{MPSb}
\end{eqnarray}
And the relative equilateral bispectrum error $E^R_{B^{eq}_{\kappa}}$ is given by :
\begin{eqnarray}
E^R_{B^{eq}_{\kappa}}=\frac{1}{N_m} \sum_m \left( \frac{B^{eq}_{\kappa^m} - B^{eq}_{\tilde{\kappa}^m} }{B^{eq}_{\kappa^m} }\right),
\label{RPSE}
\end{eqnarray}

The maximum discrepancy is obtained with the CFHTLS mask where the relative bispectrum error is about $3\%$ while is about $1\%$ with Subaru mask, which remains consistent with the sample variance in blue - dashed line on the right. This result is satisfactory because no constraint on the solution has been used to improve the estimation quality of the bispectrum.

\subsection*{Computing time:  two-point correlation versus  power spectrum using the iterative reconstruction}
As in the previous section, we have compared the computing time of the two-point correlation function applied on the incomplete shear maps with the power spectrum applied on inpainted mass map obtained from incomplete shear maps. The time to process the MCA-inpainting starting from shear maps followed by the power spectrum is 6 minutes still using a 2.5 GHz PC-linux processor. It is a bit longer than starting from convergence maps because it requires a conversion from shear field to convergence $\kappa$ field and vice versa at each iteration in the MCA-inpainting and also because the algorithm is this time written in IDL language. It still remains 80 times faster than the two-point correlation function. Furthermore, the reconstructed map can also be used to compute higher-order statistics.

\subsection*{Computing time : three-point correlation versus  inpainting reconstruction followed by a bispectrum}
Here, we compare the time needed to compute on the one hand the three-point correlation function applied on the incomplete mass maps and on the other hand, the bispectrum applied on inpainted masked shear maps. Our three point-correlation function estimator is based on the  averaging of the power of the equilateral triangles on the direct space and it is not biased by missing data. But its computational time is very long. Even using the equilateral information, more than 8 hours are needed to process the three-point correlation function in a field of 4 square degree on a 2.5 GHz processor PC-linux.

The time to compute the equilateral bispectrum including the MCA-inpainting method all written in IDL, in the same field still using the same 2.5 GHz processor PC-linux is only 6 minutes. This is 80 times faster than the three-point correlation function. And it only requires $O(N \log N)$ operations compared to the three-point correlation estimation that requires $O(N^2)$ operations.

\section{Conclusion}
\label{sect_ccl}
This paper addresses the problem of statistical analysis of Weak Lensing surveys in the case of incomplete data.

We have presented a method for image interpolation across masked regions and its applications to weak lensing data analysis. The proposed inpainting approach relies strongly on the ideas of MCA \citep{inpainting:starck04} and on the sparsity of the weak lensing signal in a given dictionary. The proposed reconstruction algorithm is based on a decomposition in a basis of cosines that turned out to be the best representation for weak lensing data. This recent tool can be applied to many other interesting applications and it is already used in CMB data analysis to fill in the galactic region \citep{inpainting:abrial06}. 
The algorithm has been extended to the weak lensing inversion problem with realistic masks. With this extension, the inpainting method provides a solution for the problem of estimation of the convergence mass map from incomplete shear maps.

We have proposed to use our fast $O(N \log N)$ inpainting algorithm to lower the impact of missing data on statistics estimations. We have shown that our inpainting method enables us to use the power spectrum in future large weak lensing surveys by filling in the gaps in weak lensing mass maps in the presence of noise. We have shown that our inpainting method enables to reach an accuracy of about 1\% with the CFHTLS mask and about 0.3\% with Subaru mask for the power spectrum 

We have shown that our inpainting technique can also be applied to higher-order statistics. In particular, we have presented a fast and accurate method to calculate the equilateral bispectrum using a polar FFT and we have shown that our inpainting method enables to reach an accuracy of about 3\% with the CFHTLS mask and about 1\% with Subaru mask for the bispectrum.

It would be interesting to extend the MRLENS filtering package \citep{starck:sta05} in order to build a filtered mass map from incomplete shear maps by applying the inpainting technique.

In future work, this inpainting technique can be extended to compute other non-gaussian statistics : higher-order statistics, peak finding, etc... taking advantage of the map recovery.

\subsubsection*{Software}
The software related to this paper, {\bf FASTLens}, and its full documentation 
is available from the following link:
\begin{verbatim}
       http://jstarck.free.fr/fastlens.html 
 \end{verbatim}

\section*{Acknowledgments}
This study has made use of simulations performed in the context of the HORIZON project. The N-body simulations used here were performed on the Grid'5000 system and deployed using the  DIET software. Therefore, this work has been partially supported by the LEGO (ANR-CICG05-11) grant. One of the authors would like to thank Emmanuel Quemener and Benjamin Depardon for porting the RAMSES application on Grid'5000. 
We wish also to thank Yassir Moudden and Joel Berge for useful discussions and comments and Pierrick Abrial for his help on computational issues.

\bibliographystyle{astron}
\bibliography{PSinpainting.bib}

\label{lastpage}

\end{document}